\documentclass[a4paper,10pt]{article}
\usepackage{psfrag}
\usepackage{graphicx}
\usepackage{xcolor}
\usepackage{afterpage}
\usepackage{amsmath}
\usepackage{amsfonts}
\usepackage{amsthm}
\usepackage{amssymb}
\usepackage{setspace}
\usepackage{subfigure}
\usepackage{enumerate}
\usepackage{natbib}
\usepackage{rotating}
\usepackage{setspace}
\usepackage{lineno}
\usepackage{url}
\usepackage[margin=3cm]{geometry}
\usepackage{caption}

\usepackage{bbm}

\graphicspath{{./IntCal20Plots/}}

\newcommand{\rC}{$^{14}$C }
\newcommand{\nC}{$^{12}$C }
\newcommand{\rCyrs}{$^{14}$C yr BP }
\DeclareMathOperator{\EX}{\mathbb{E}}
\providecommand{\keywords}[1]{\textbf{\textit{Keywords:}} #1}

\title{Non-parametric calibration of multiple related radiocarbon determinations and their calendar age summarisation}
\author{T. J. Heaton}

\begin{document}

\maketitle

\begin{abstract}
Due to fluctuations in past radiocarbon ($^{14}$C) levels, calibration is required to convert $^{14}$C determinations $X_i$ into calendar ages $\theta_i$. In many studies, we wish to calibrate a set of related samples taken from the same site or context, which have calendar ages drawn from the same shared, but unknown, density $f(\theta)$. Calibration of $X_1, \ldots, X_n$ can be improved significantly by incorporating the knowledge that the samples are related. Furthermore, summary estimates of the underlying shared $f(\theta)$ can provide valuable information on changes in population size/activity over time. Most current approaches require a parametric specification for $f(\theta)$ which is often not appropriate. We develop a rigorous non-parametric Bayesian approach using a Dirichlet process mixture model, with slice sampling to address the multimodality typical within $^{14}$C calibration. Our approach simultaneously calibrates the set of $^{14}$C determinations and provides a predictive estimate for the underlying calendar age of a future sample. We show, in a simulation study, the improvement in calendar age estimation when jointly calibrating related samples using our approach, compared with calibration of each $^{14}$C determination independently. We also illustrate the use of the predictive calendar age estimate to provide insight on activity levels over time using three real-life case studies.
\end{abstract}

\section{Introduction}
Since its development by Willard Libby and colleagues \citep{Anderson1947, Libby1949, Arnold1949}, radiocarbon ($^{14}$C) dating has revolutionised archaeological and environmental science. Radiocarbon dating relies on the simple idea that, while alive, organisms take in carbon from their surroundings and so have a ratio of \rC to \nC  that is in equilibrium with their atmosphere. Once an organism dies it stops taking in new carbon, the stable \nC remains but the level of \rC halves every 5730 years. Measurement of the ratio of \rC to \nC left within a sample therefore enables a dating technique which can extend back 55,000 years --- samples from further back in time have so little \rC remaining that they cannot be reliably measured. If the concentration of atmospheric \rC had been constant over this time, \rC dating would be straightforward. However, it has fluctuated significantly. To date samples precisely, scientists need to \textit{calibrate} their \rC determinations against a record of past \rC levels to transform them into calendar ages. Without this calibration, \rC determinations are not directly interpretable. 

\begin{figure}
\centering
\includegraphics[width = 0.8\textwidth, keepaspectratio]{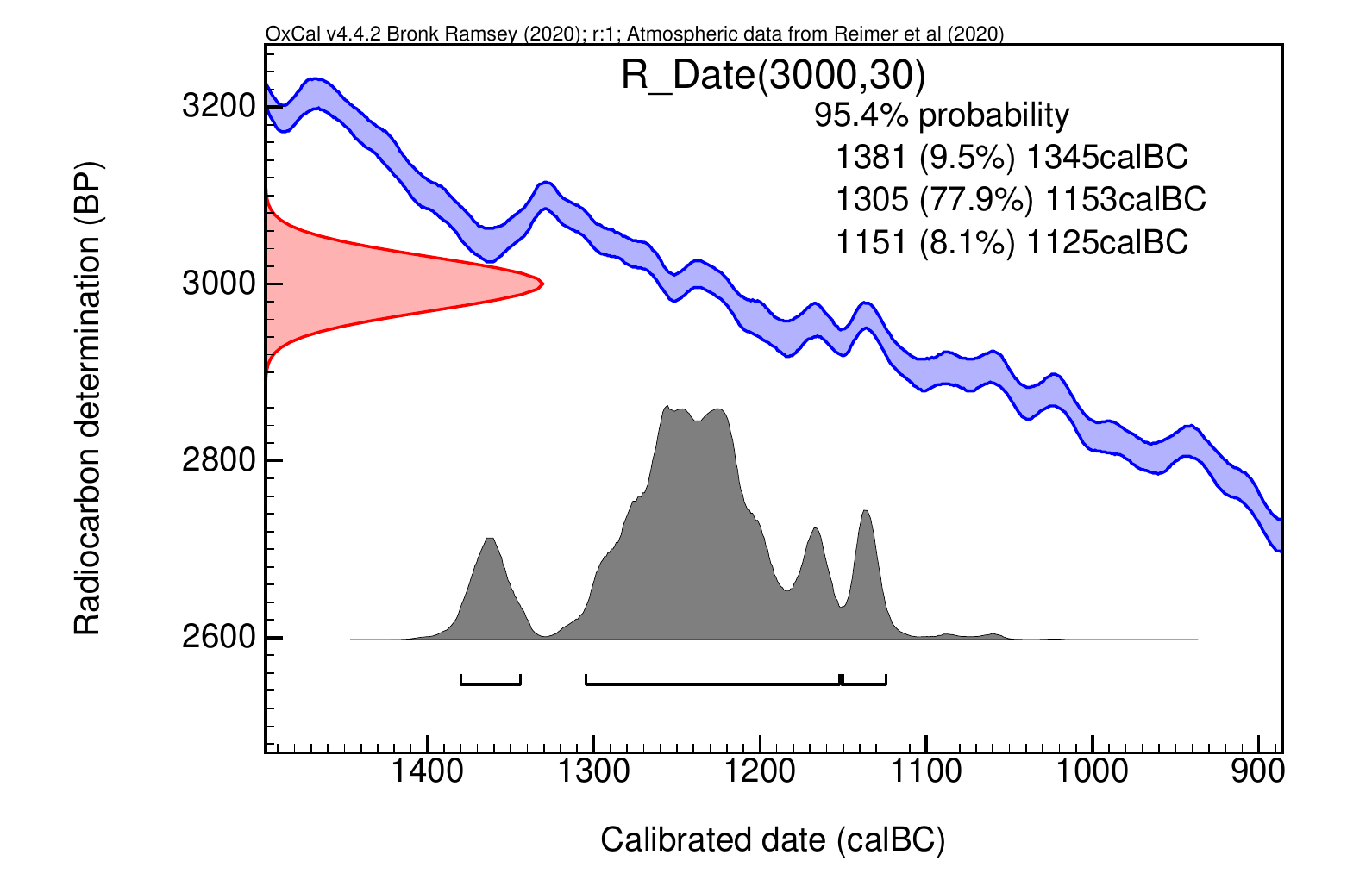}
\caption{\textit{OxCal calibration of an object with a radiocarbon determination of $3000 \pm 30$ $^{14}$C yr BP. The observed \rC determination was modelled by $X_i \sim N(\mu(\theta_i), 30^2)$. The $y$-axis shows the radiocarbon determination; the current IntCal20 \citep{Reimer2020} calibration curve together with its 1-sigma (i.e., 68\%) probability intervals is shown in blue; and along the $x$-axis we show the posterior estimate for the object's calendar age $\theta_i$ (shown here  as BC). The 95.4\% highest probability density intervals for the sample's calendar age are given in the top right hand. Intuitively, the posterior calendar age estimate consists of those dates for which the calibration curve is consistent with the observed $X_i$.} \label{F:OxCalExample}}
\end{figure}

In this paper, we consider the problem of calibrating and summarising a set of samples, with radiocarbon determinations $X_1, \ldots, X_n$, which are known to be related to one another (for example, arising from a particular site, or set of sites, populated by a particular culture). Each sample has an unknown calendar age $\theta_i$ but, since they are related, these calendar ages are assumed to arise from the same unknown prior density $f(\theta)$. This unknown density $f(\theta)$ may be related to the size of the population under study, or their activity level, at the site/sites. Specifically, we model,
\begin{align*}
\theta_1, \ldots, \theta_n &\sim f(\theta) \\
\textrm{and }X_i &\sim N(\mu(\theta_i), \sigma_i^2) \quad \textrm{for } i = 1, \ldots, n.
\end{align*}
Here $\mu(\theta)$ denotes what is known as the radiocarbon calibration curve and records the atmospheric \rC level at time $\theta$ \citep{Suess1968}; and $\sigma_i$ is the measurement uncertainty on our \rC determination $X_i$ \citep{Scott2007}. Globally-ratified estimates of $\mu(\theta)$ are provided by the IntCal working group and, as is standard in \rC dating, are not updated during calibration of the determinations $X_1, \ldots, X_n$. Figure \ref{F:OxCalExample} illustrates the typical calibration of a single \rC determination $X_i$ using the IntCal20 calibration curve \citep{Reimer2020}. See Section \ref{S:Calibration} for more details on calibration.

Our aim is two-fold. Firstly, we desire to calibrate the set of \rC determinations and provide a posterior estimate of each calendar age $\theta_i | X_1, \ldots X_n$. Here, we wish to use all the determinations $X_1, \ldots, X_n$ as they provide information on the shared underlying density $f(\theta)$ from which the $\theta_i$ are drawn. Typically we would expect incorporating such information would lead to improved accuracy in our posterior estimates of the $\theta_i$, as we explain further in Section \ref{S:MultipleDets}.

Secondly, in addition to the calibration of individual samples, many users are interested in summarising the calendar age information provided by a set of \rC determinations: to obtain a proxy for population size or activity levels over time. In this context, one is more interested in estimating the underlying density $f(\theta)$ given the \rC determinations we have observed. We therefore also aim to provide such an estimate alongside the calibration of the individual samples. Since we work in the Bayesian setting, this will be in the form of a predictive distribution for the calendar age of a future hypothetical object from the same site/set of sites , i.e., $f(\theta_{n+1} | x_1, \ldots, x_n)$. Periods of time when we predict a higher density of objects may relate to times when the underlying culture had a higher level of activity or was more numerous; and conversely periods when there are fewer objects may indicate times when the culture was less successful or smaller.

Significant recent increases in the availability and use of \rC determinations make these questions particularly timely. Advances in \rC measurement techniques \citep{Synal2007,Suter99} have enabled much smaller samples to be dated. This has increased the number of potentially dateable samples on any archaeological site into the thousands, all of which may provide crucial historical understanding as to the use of that site over time \citep{Bayliss2009}.

The majority of previous methodological research into the calibration and summarisation of multiple related \rC determinations has been parametric. These \textit{phase} models require strong, a priori, assumptions as to the parametric form of the underlying calendar age density $f(\theta)$ --- for example, uniform \citep{Buck92}, triangular \citep{BronkRamsey09} and trapezoidal \citep{Karlsberg2006, LeeRamsey12}. However, in many cases it is either not possible or not desirable to specify the form of the calendar age density so rigidly in advance and a non-parametric approach is needed. 

We are not aware of a statistically-rigorous non-parametric approach to the joint calibration and summarisation of multiple related \rC determinations. Currently, most users, if they do not wish to specify a specific parametric phase model, calibrate multiple \rC determinations as though they were independent even when they know their samples are related, i.e., they consider only $\theta_i | X_i$ for each sample separately and without consideration of the shared underlying density $f(\theta)$. This is expected to give suboptimal calendar age estimation. Incorporating information on $f(\theta)$, which can be obtained from the other \rC determinations, should improve calibration accuracy. 

Summarisation of the calendar ages of the multiple samples, i.e., provision of an estimate of $f(\theta)$, is typically done as a later and separate step with most users creating summed probability distributions (SPDs, also called summed probability functions, SPFs; summed calibrated probability distribution, SCPDs; or cumulative probability density functions, CPDFs). These are simply created by adding together the independent posterior density estimates for each $\theta_i | X_i$. They are neither statistically sound as a method of providing a predictive age estimate, nor provide an indication of the uncertainties in the resultant estimate which makes their interpretation challenging --- see \citet{Williams12} and \citet{Contreras2014} for reviews. Equivalent approaches to SPDs are also used for other dating methods, notably fission-track and optically stimulated luminescence (OSL) dating. In these fields, SPD equivalents have been significantly criticised as failing to properly recover two components of a mixture distribution \citep{Galbraith1988, Galbraith1998B, Galbraith2010}.

\citet{Ramsey2017} has proposed an iterative approach to joint calibration and summarisation. Each determination is first calibrated individually and a value is drawn from the posterior of $\theta_i | X_i$ for $i = 1, \ldots, n$. Next, a kernel density estimate is fitted to these $n$ specific drawn values to generate an initial estimate of the shared age density $f(\theta)$. This kernel estimate is then reused as a prior for each $\theta_i$ to update the posterior $\theta_i | X_i$. This process is repeated until convergence is reached. While such a heuristic approach offers significant improvements over SPDs and independent calibration, it lacks some elements of formal statistical underpinning being composed of both frequentist and Bayesian elements.   

Here, we present a fully Bayesian method, fitting within the Bayesian paradigm used universally for radiocarbon calibration, that aims to provide both improved calendar age estimation of the individual samples, and a statistically-rigorous predictive density estimate for a new object. We employ a Bayesian non-parametric approach through an infinite Dirichlet Process Mixture Model \citep[DPMM,][]{Neal2000, Walker2007} in combination with slice sampling \citep{Neal2003} to perform the calibration. Intuitively, this DPMM assumes our sampled objects arise from an unknown number of archaeological clusters. Calendar ages $\theta_i$ of the samples are estimated jointly at each step recognising they arise from the unknown DPMM density $f(\theta)$ with the consequence that their estimation will hopefully be improved. The predictive density of the calendar age of a new object is also obtained, along with pointwise credible intervals, to allow practical interpretation of potential changes in historical activity over time. 

Through the incorporation of slice sampling, we are able to implement our approach almost entirely via direct sampling from the conditionals within a Gibbs sampler, making estimation fast. Furthermore, the slice sampling updates to the calendar ages $\theta_i | X_1, \ldots X_n$ are ideally suited to address the multimodality in these estimates which occurs as a result of the inherent non-monotonicity in the calibration curve $\mu(\theta)$. To update our DPMM we investigate and compare two different approaches --- a P\'olya urn approach \citep{Neal2000} and a quicker approach which also uses slice sampling \citep{Walker2007}. Comparisons of these two approaches to DPMM sampling have previously been predominantly restricted to small and synthetic examples \citep{Hastie2015}.   

Our paper is laid out as follows. In Section \ref{S:RadiocarbonIntro} we provide a brief introduction to \rC dating and calibration. Section \ref{S:MultipleDets} sets out our specific questions regarding the optimal calibration and summarisation of multiple \rC determinations, and reviews approaches previously taken in the literature --- both parametric and non-parametric. In particular, we explain why SPDs, currently the most popular non-parametric approach to summarising multiple \rC determinations, are not a statistically rigorous or suitable method. In Section \ref{S:NPBayes} we present our alternative non-parametric Bayes approach, detailing how a DPMM and calibration can be combined. Section \ref{S:SimStudy} provides a simulation study demonstrating how our approach, with joint calibration of the $X_i$ \rC determinations, offers significant improvements in estimation of the calendar ages $\theta_i$ compared to independent calibration. We also investigate the estimates of the underlying shared density $f(\theta)$ obtained via our approach and SPDs; and compare the two approaches (P\'olya Urn and slice sampling) to the updating of the DPMM component in our model. Section \ref{S:RealExamples} presents three real-life examples of the summarisation of multiple \rC determinations and the insight, in terms of the level of archaeological activity over time, which can be provided by estimation of the underlying calendar age distribution $f(\theta)$. These practical examples consider the prevalence, over time, of Irish medieval settlements known as raths \citep{Kerr2014}; whether climate change may have caused a population decline in Ireland at the end of the European Bronze Age \citep{Armit2014}; and changes in North American palaeoindian demography over the Younger-Dryas and into the Holocene \citep{Buchanan2008}. We show the density estimates for the underlying $f(\theta)$ provided by both SPDs and our non-parametric Bayes approach. Finally, in Section \ref{S:Conclusion} we summarise our work and provide suggestions for further study. The current development version of R code, including all examples and the simulation study, is accessible via GitHub (\url{https://github.com/TJHeaton/NonparametricCalibration}).

\paragraph{Notation} As standard in the radiocarbon literature, all ages in this paper are reported relative to mid-1950 AD (= 0 BP, before present). The pre-calibration \rC ages/determinations, $X_i$, are given in units ``\textit{$^{14}$C yr BP}''. Calendar (or calibrated) ages, $\theta_i$, are denoted as ``\textit{cal yr BP}''; or sometimes ``AD" when the estimated calendar ages are recent. 

\section{Radiocarbon dating and the need for calibration\label{S:RadiocarbonIntro}}

\subsection{Radiocarbon dating}
The original approach to radiocarbon dating \citep{Libby1949} relied upon an assumption that the ratio of \rC to $^{12}$C in the atmosphere had been constant throughout time. Under such an assumption, after accounting for isotopic fractionation, all samples would have had the same isotopic (\rC to $^{12}$C) ratio at the point they stopped interacting with their local environment no matter at what point in the past that was. Consequently, given any sample, one can determine a \textit{radiocarbon age}, $X_i$, according to radioactive decay on the basis it had begun with the specific \rC to $^{12}$C ratio of a standard \citep{stuiver_polach_1977}. 

However, it was soon discovered \citep{deVries1958, Willis1960} that the ratio of \rC to $^{12}$C had varied considerably over time and radiocarbon ages did not precisely correspond to the calendar ages of the samples. These \rC variations are due to a range of factors such as fluctuations in solar activity and geomagnetic field strengths that modify impinging cosmic radiation levels; and changes to the carbon cycle which can release large stores of very old carbon deficient in $^{14}$C. To improve the accuracy of radiocarbon dating we therefore need to adjust or calibrate our radiocarbon ages/determinations $X_i$ to provide calendar ages $\theta_i$.

Determining the historic proportion of \rC in the atmosphere, and hence improving radiocarbon dating, is done via the collection of reference historic objects in which we can both directly measure $^{14}$C and obtain an independent estimate of their calendar age.  Measurement of tree-rings currently enable us to create a record of atmospheric \rC back to about 13,900 years from the present. Further back in time, other objects are used such as corals; stalagmites; and foraminifera and macrofossils found in ocean and lake sediments --- all of which contain atmospheric \rC and can be approximately dated using other means \citep{Reimer2020}.

\subsection{A calibration curve}
Given a set of reference objects for which we have both \rC measurements and independently known (or estimated) calendar ages, we can create what is known as a \textit{calibration curve}. This calibration curve $\mu(\theta)$ is a mapping providing, for an object of true calendar age $\theta$, the corresponding radiocarbon age. Given a undated object for which we obtain a \rC determination, $X_i$, one can estimate its calendar age by inverting this mapping as described in Section \ref{S:Calibration}. Typically, $\mu(\theta)$ is highly non-monotonic due to the variations in past \rC levels, see Figure \ref{F:OxCalExample}. Multiple calendar ages could therefore correspond to the same radiocarbon determination.   

The first atmospheric calibration curve for the Northern Hemisphere was produced by \citet{Suess1968}. Since then much work has been done on improving these curve estimates and extending them further back in time. The modern, internationally-ratified, standard for the radiocarbon calibration curve is known as IntCal and is updated regularly as new datasets become available. The first IntCal curve was generated in 1998 (IntCal98) and has been updated in 2004, 2009, 2013 and 2020. IntCal20 \citep{Reimer2020} is the current version agreed for use by the community --- accompanying curves are also provided for the Southern Hemisphere \citep{SHCal20} and the surface oceans \citep{Marine20}.

Since 2004, the IntCal curves have been estimated in a Bayesian framework \citep[see][for details]{Buck2004, Blackwell2008, Heaton2009, Niu2013, HeatonIntCalStats2020}. Construction of the most recent IntCal20 curve \citep{HeatonIntCalStats2020} is performed via Bayesian splines with errors-in-variables \citep{Berry2002}. Given the reference calibration data, IntCal20 provides pointwise estimates of the posterior mean and standard deviation of $\mu(\theta)$, the radiocarbon age for an object of calendar age $\theta$, on a regular grid back to 55,000 cal yr BP.

\subsection{Calibration of an object of unknown age \label{S:Calibration}}
To calibrate an object of unknown age, i.e., provide an estimate of its calendar date $\theta_i$, one compares its observed radiocarbon determination $X_i$ against the calibration curve. Since this estimation of $\theta_i$ requires inversion of $\mu(\cdot)$, calibration is also performed using a Bayesian approach. Specifically, we model
\begin{align*}
X_i &= \mu(\theta_i) + \epsilon_i,
\end{align*}
where $\theta_i$ is the unknown calendar age we wish to estimate; $\mu(\theta_i)$ is the value of the calibration curve at time $\theta_i$; and $\epsilon_i \sim N(0, \sigma_i^2)$ the error in the measurement of the sample's radiocarbon. To complete the model we specify our prior on both $\mu(\cdot)$  and $\theta_i$:
\begin{align*}
\mu(\theta) | \theta &\sim N(m(\theta), \rho(\theta)^2), \textrm{ and} \\
\theta_i &\sim f(\theta),
\end{align*}
where $m(\theta)$ and $\rho(\theta)$ are the pointwise calibration curve posterior mean and standard deviation values provided by IntCal --- these values are treated as known during calibration; and $f(\theta)$ contains any prior information we might have on the object's calendar age. Note that we can marginalise over the calibration curve so that, $X_i|\theta_i \sim N(m(\theta_i), \rho(\theta_i)^2 + \sigma_i^2)$.

By far the most popular software used by the radiocarbon community to perform calibration is OxCal \citep{BronkRamsey09} which currently implements the above estimation of $\theta_i$ via Metropolis-Hastings. In Figure \ref{F:OxCalExample} we present the calibration of a single radiocarbon determination of $X_1 = 3000$ \rCyrs (with $\sigma_1 = 30$) against the IntCal20 curve within the OxCal software (version 4.3). The resultant posterior calendar age estimate $\theta_1$ is shown along the bottom. Note in particular the non-monotonic nature of the calibration curve $\mu(\theta)$, this is typical and a result of the discussed historic fluctuations in \rC production. Multi-modal calendar age estimates of individual determinations, $\theta_i$, such as seen here, are therefore extremely common in \rC calibration and can cause difficulty in mixing for more complex Bayesian calibration problems. We aim to overcome this additional challenge through our use of slice sampling.

\section{Calibrating and summarising related \rC determinations \label{S:MultipleDets}}
\subsection{Statistical Formulation \label{S:Formulation}}
In this paper, we consider the calibration, and simultaneous calendar age summarisation, of multiple \rC determinations $X_1, \ldots, X_n$ arising from a set of objects that are believed to have calendar ages which are related to one another. Such relations might occur when sampling objects that arise from the same site, group of peoples, culture, or type of artefact. We wish to use the knowledge that the underlying calendar ages $\theta_i$ are related to improve our calibration, and to provide an useful summary of the combined (calibrated) calendar age information provided by the set of objects.  Specifically, we assume each unknown $\theta_i$ arises from the same underlying, but unknown, calendar age distribution $f(\theta)$, so that
\begin{align*}
\theta_1, \ldots, \theta_n &\sim f(\theta), \textrm{ and} \\
X_i &\sim N(\mu(\theta_i), \sigma_i^2) \quad \textrm{for } i = 1, \ldots, n.
\end{align*}

Knowledge that the unknown calendar ages of the objects are drawn from the same shared distribution, even if that shared distribution is unknown, should mean that we can improve our calibration by borrowing information from the other objects. It is well accepted that independent calibration of multiple radiocarbon determinations, i.e., $\theta_i|X_i$, provides calendar age estimates that are more spread out than their true calendar ages \citep[see][]{Buck92, Ramsey2017}. We expect to obtain better calendar age estimation by considering $\theta_i | X_1, \ldots, X_n$, which will tend to shrink the posterior calendar age estimates towards one another.

In our Bayesian framework, this is done by considering the unknown calendar ages $\theta_1, \ldots, \theta_n$ to be drawn from some larger population on which we place a prior $f(\theta)$. Given $f(\theta)$, each $\theta_i$ is then conditionally independent of $X_j$ for all $i \neq j$. Even when the form of $f(\theta)$ is unknown, the underlying shared distributional assumption should still be used for calibration.

Through our approach, we are able to estimate $f(\theta)$ using the information given by $X_1, \ldots, X_n$ simultaneously to the calibration process. This significantly improves estimation of the individual calendar ages as shown in Section \ref{S:SimStudy}. Furthermore, the resultant predictive distribution for the calendar age of a future hypothetical object, i.e., $f(\theta_{n+1} | X_1, \ldots, X_n)$ can also be used as a proxy to provide practical archaeological insight into the varying use of the site, activity of the peoples, culture, or type of artefact over time.  Indeed it is often this summary calendar age distribution which is of most interest to \rC users, see Sections \ref{S:DensityRecon} and \ref{S:RealExamples} for details.

\subsection{Current Methods to Calibrate and Summarise \rC Determinations  \label{S:CurrentApproaches}}

\subsubsection{Parametric Approaches}
If we are willing to select, a priori, a specific parametric form for the shared calendar age density $f(\theta)$, for example
$\theta_1, \ldots, \theta_n \sim f(\theta) = N(\phi, \tau^{-1})$, then it is straightforward to provide both a posterior for the underlying calendar age distribution's parameters (e.g., the mean $\phi$ and precision $\tau$ in the case of the normal density above); and each individual $\theta_i$ \citep[for more details see][]{Naylor88, Buck92}. Implementations for various underlying families of calendar age densities have been proposed, e.g., a uniform density with unknown start and end dates \citep{Buck92, Christen94}, normal, triangular \citep{BronkRamsey09}, and trapezoidal \citep{Karlsberg2006, LeeRamsey12, RamseyLee13}. Such parametric approaches are called \textit{phase} models as they assume the site has gone through a phase of use. Estimation of $f(\theta)$ for fixed mixtures of normal phases, but without simultaneous calibration of the \rC determinations, is also possible in BCHRON\citep{BCHRON} and STAN \citep{Price2020}.

However, in many cases it is not possible/desirable to specify the form of $f(\theta)$ in advance. It may not be appropriate to model the usage of the site by a simple parametric form. Alternatively, objects from the site may come from a mixture of different time periods with an unknown number of phases of high activity interspersed with little/no activity as the site falls into/out of development over time. Instead we may prefer a non-parametric approach letting the data themselves inform us as to the shape of the density or the number of distinct periods.

\subsubsection{Non-Parametric Approaches \label{S:SPDExplain}}

\paragraph{Calibration of Multiple Determinations}
When presented with a set of \rC determinations $X_1, \ldots, X_n$ for which a parametric phase model is deemed inappropriate, the majority of \rC users will calibrate each $X_i$ independently from the others with an uninformative prior on $\theta_i$, i.e., calculate
\begin{equation} \label{Eq:IndCalib}
f_i(\theta_i | X_i ) \propto f(X_i | \mu(\theta_i)) f(\mu(\theta_i) | \theta_i) \quad \textrm{ for } i = 1, \ldots, n,
\end{equation}
even if they know the underlying $\theta_i$ calendar ages are related and are drawn from a shared density $f(\theta)$. Such an approach neither includes estimation of the joint $f(\theta)$, nor allows one to share information across $X_1, \ldots, X_n$. As explained in Section \ref{S:Formulation}, and as we show in Section \ref{S:SimStudy}, this leads to suboptimal calendar age estimation.  

\paragraph{Summed Probability Distributions}
The most common approach to summarise calendar age information of multiple \rC determinations is via summed probability distributions \citep[SPDs, see][]{Williams12, Contreras2014}. Here, the posterior calendar age density $f_i(\theta_i | x_i)$ of each object is first calculated independently from the others as in (\ref{Eq:IndCalib}). These individual densities are then summed/averaged to give an SPD estimate,
$$
f(\theta | x_1, \ldots, x_n) = \frac{1}{n} \sum f_i(\theta_i | x_i).
$$
Summed probability distributions are not statistically valid estimators of the calendar age of a potential future object. The independence assumed in the separate calibration of each $X_i$, followed by subsequent summarisation, generates a contradiction. Either each individual $\theta_i$ is differently distributed, in which case separate calibration is legitimate, but it is not then appropriate to jointly summarise the resultant calendar ages; or alternatively the objects are related and a summary is appropriate, but in which case the objects should not be calibrated separately from one another. Additionally, SPDs are not a predictive density, they simply provide an estimate of the calendar age were you to resample an object at random from the set of $n$ objects on which they are based. An SPD implicitly assumes this sample of $n$ objects provides the exhaustive set of possible calendar ages, i.e., that there are no more possible objects one could ever date. This is not what we desire. Our $n$ observed objects (and their calendar ages) will typically form a small sample drawn randomly from a much wider population of potential objects (and calendar ages). In summarising, our aim is to make inference about this wider population. 

Furthermore, SPDs do not come with uncertainties on the summary estimate. These are key if one wishes to make inference on activity levels over time. The non-monotonicity of the calibration curve $\mu(\theta)$ means that each $\theta_i | X_i$ can be multimodal, one must therefore be careful in interpreting each peak in a summarised density as indicating a separate period of activity. For example, in Figure \ref{F:OxCalExample}, we see our single determination $X_1$ results in a highly multimodal calibrated calendar age estimate. However, since we only have a single object it would be false to infer that this is evidence for multiple distinct periods of activity (e.g., around BC 1360, 1230 and 1140) separated by periods of inactivity. We therefore need credible intervals on our summarised density estimate. To compound this problem, the independent calibration of the objects within an SPD tends to create many small peaks within the resultant summary. In particular, a single high-precision \rC determination can create a sharp peak in the SPD and completely ruin the estimate of $f(\theta)$. Our proposed Bayesian approach provides credible intervals for the summary estimate $f(\theta)$; while also reducing the number of spurious small peaks and troughs due to minor calibration curve inversions, as shown in Section \ref{S:RealExamples}.

Several alternatives and adjustments to SPDs have been suggested, however all those of which we are aware still suffer from some of the weaknesses above. \citet{Kerr2014} propose that the risk of over-interpreting SPDs due to calibration curve artefacts can be reduced if the initial SPD is binned to provide coarse histograms. However this appears an ad hoc approach lacking a theoretical justification. Furthermore, they provide no guidance on how coarse a binning to apply. Too wide a bin would lead to the loss of significant real information, too narrow would not remove the curve artefacts. An equivalent problem to summarisation has also been considered by \citet{Dye2016} through tempo-plots. A tempo plot aims to provide, given a set of radiocarbon determinations $X_1, \ldots, X_n$, an estimate $N(\theta)$ of the number of those observations occurring before a time $\theta$ with the intention this can be interpreted as a proxy for overall historical activity. In our context of density estimation, the tempo-plot's construction make it mathematically identical to the distribution function one would obtain by integrating the SPD formed from the underlying determinations. The tempo-plot's credible intervals for $N(\theta)$ are then the confidence intervals of the sum $\sum_{i=1}^n Ber(p_i)$ of independent Bernoulli random variables with probabilities set to the individual values of each object's age distribution functions $p_i = F_i(\theta)$. Approaches to perform hypothesis testing of, e.g., $H_0: f(\theta) = f_0(\theta)$ using Monte Carlo methods on the SPD have also been suggested \citep{Shennan2013, Edinborough2017} although these are incongruous, following a Bayesian calibration step, and incoherent SPD summarisation, with a classical hypothesis test.

\paragraph{Alternative approaches to multiple calibration and summarisation} 
\citet{Ramsey2017} recently proposed a method combining the joint calibration of multiple \rC determinations, under the assumption their calendar ages arise from a shared non-parametric distribution $f(\theta)$, with kernel density estimation to estimate $f(\theta)$. This iterative approach alternates between a step which calibrates each determination conditional on the current estimate for $f(\theta)$, i.e., $\theta_i | X_i, f$; and a step updating the estimate of $f(\theta)$ given $\theta_1,\ldots, \theta_n$. This latter step is performed by sampling from $\theta_i | X_i, f$ for $i = 1, \ldots, n$ and then fitting a kernel density estimate to the sampled values. The process of calibration under the current estimate for the prior $f(\theta)$, and then kernel density updating of $f(\theta)$ is repeated until convergence is obtained. This method of \citet{Ramsey2017} is similar in principle to our proposed approach, except that the kernel density update introduces a frequentist element to an otherwise Bayesian method, making the overall method somewhat of a hybrid. Our DPMM approach adds statistical consistency and rigour to provide a fully Bayesian approach to the non-parametric calibration and summarisation of multiple \rC determinations.

\subsection{Other considerations when summarising \rC datasets}
It is necessary to consider several potentially complicating effects in interpreting the estimated summary density $f(\theta)$ as a proxy for overall activity, see \citet{Williams12} and \citet{Contreras2014}. Taphonomic loss (i.e., a reduction in sampling of older objects at a site due to increased likelihood of site destruction through time) and non-random sampling (e.g., where archaeologists tend to concentrate excavations on time periods of particular historic interest and are more likely to have sampled objects from these periods) are common in archaeological studies. We do not specifically address these issues in this work. Taphonomic loss might be accounted for through the introduction of a (known) function $g(\theta)$ giving the probability that a sample of calendar age $\theta$ will be preserved for the modern day. Using this taphonomic preservation probability $g(\theta)$, we might then either adjust our observational model to account for the fact samples are lost; or, more simply, consider a post-estimation transformation of the predictive density subject to taphonomic loss to one without. Non-random sampling is significantly more difficult to address without knowledge of its structure. In cases where this is likely we therefore urge caution in interpretation. Note that these issues are only of concern when interpreting the summary density estimate. They do not affect the validity, or benefit, of using the multiple samples jointly to improve calibration of the samples. When improving estimation of the calendar ages of the individual objects, we are concerned with the specific population of objects one might sample --- this does not need to be representative of wider activity.   

The non-monotonic nature of the calibration curve $\mu(\theta)$ can also raise identifiability issues in calendar age estimation and summarisation that are partly intractable. In particular, when the range of calendar ages of a set of objects is narrow and corresponds to a significant wiggle in the calibration curve, there may be multiple calendar periods which could equally generate the given \rC determinations. Here, the summarised density estimate $\hat{f}(\theta)$ is likely to place some mass on each time period generating multiple peaks even if the samples only arose from one. This is discussed in detail in Section \ref{S:DensityRecon}. We recommend all \rC determinations (and the summarised calendar age estimates) are initially plotted alongside the calibration curve to assess this possibility and amend the intepretation accordingly. Such plots are provided for all our examples in Sections \ref{S:SimStudy} and \ref{S:RealExamples}. The credible intervals we provide for $\hat{f}(\theta)$ should also help assessment.

\section{Bayesian non-parametric calibration and summarisation \label{S:NPBayes}}

Our approach alternates between updates of the calendar ages $\theta_i$, given $X_i$ and $f(\theta)$; and updates to $f(\theta)$, given $\theta_i$ for $i=1, \ldots, n$. These two steps are combined within a Gibbs MCMC scheme. The sampler proceeds by direct sampling from the relevant conditionals in almost all its updates. Consequently, the overall MCMC mixes rapidly and the overall method is quick to implement. 

We create a non-parametric prior on the underlying calendar age distribution $f(\theta)$ via a Dirichlet Process Mixture Model \citep[DPMM, e.g.,][] {Neal2000}. Given the calendar ages $\theta_i$, this DPMM can be updated by direct sampling (with the exception of a single hyperparameter updated by Metropolis-Hastings). We consider two different approaches to this updating of the DPMM component. The first uses a P\'olya urn approach as proposed by \citet{Neal2000} where the mixture weights are integrated out. Each calendar age is specifically labelled (allocated) as belonging to a particular component of the infinite mixture. These allocations are then updated, one-at-a-time and conditional upon all the other allocations, within the sampler. This approach to label updating potentially makes mixing more difficult and slows convergence. The second approach uses a slice sampling approach suggested by \citet{Walker2007}. This approach retains the allocations but does not integrate out the mixture weights and is able to sample directly from the full conditionals of the stick-breaking DPMM via a Gibbs sampler. We compare the two approaches to DPMM updating in Section \ref{S:SimStudy}.          

To update the calendar ages $\theta_i$ for each object, given the current estimate for $f(\theta)$, we perform calibration of the observed $X_i$ using the DPMM as the prior on $\theta_i$. The specifics of our DPMM updating, whereby each object is allocated to a specific component of the infinite mixture, mean that in this calibration step we do not need to work with the full prior $f(\theta)$. Conditional on the allocation, the prior on $\theta_i$ is reduced to a simpler $\theta_i \sim N(\phi_{c_i}, \tau_{c_i}^{-1})$, where $c_i$ is the component to which object $i$ belongs, see Section \ref{S:Model} for details. This calibration is done using a different slice sampling approach \citep{Neal2003} which allows direct sampling of $\theta_i | X_i, f(\theta)$. This conditional will typically be multimodal due to the non-monotonicity of the radiocarbon calibration curve --- as discussed in Section \ref{S:RadiocarbonIntro}, and illustrated in Figure \ref{F:OxCalExample}. The slice sampling aims to permit easier sampling from this $\theta_i | X_i, f(\theta)$, which feeds back into the DPMM, and hence should improve mixing of the overall MCMC. 

\subsection{Intuitive Explanation of the Model}
We model our unknown calendar age density $f(\theta)$ as an infinite mixture of individual components, or clusters. In our case, these individual calendar age components will be normal densities that can have different locations and spreads. In some cases, this mix of normal densities may represent true and distinct underlying normal archaeological phases, in which case additional practical inference may be possible. However this is not required for the method to provide good estimation. 

Each object is then considered to be drawn from one of the (infinite) clusters which constitute the overall $f(\theta)$. The probability that it is drawn from a particular cluster will depend upon the relative weight given to that specific cluster. It will be more likely than an object will come from some clusters than others. Given an object belongs to a particular cluster, its prior calendar age will be normally distributed with the mean and variance of that cluster. The mean and variance of each individual normal cluster that constitutes the overall $f(\theta)$, together with the weightings associated to each cluster, will be estimated based upon the set of \rC determinations $X_1, \ldots, X_n$ we observe. Our model is thus built as follows:
\begin{itemize}
 \item Each object has a \rC determination $X_i \sim N(\mu(\theta_i), \sigma_i^2)$. Here, $\theta_i$ is the unknown calendar age of the object we wish to estimate; $\mu(\cdot)$ is the calibration curve, provided as pointwise means and variances by the IntCal curves \citep{Reimer2020}; and $\sigma_i$ is the measurement uncertainty reported by the laboratory performing the measurement
 \item Each object belongs to a specific (calendar age) cluster of objects identified by $c_i$, i.e., object $i$ belongs to cluster $c_i$.
 \item If an object belongs to cluster $j$, i.e., if $c_i = j$, then we have a prior on its calendar age $\theta_i |\, c_i = j \sim N(\phi_{j}, \tau_{j}^{-1})$. The mean $\phi_{j}$ and precision $\tau_{j}$ of each cluster are unknown. 
\item The probability that an object belongs to cluster $j$, i.e., $P(c_i = j)$, will depend upon $j$. We place a \textit{stick-breaking} prior on the weight of each cluster (i.e., the probability an observation belongs to cluster $j$) which we then adaptively update within our DPMM.   
\item The mean $\phi_{j}$ and precision $\tau_{j}$ of each individual calendar age cluster vary according to the cluster $j$. Their values are themselves drawn from a prior distribution.
\end{itemize}

\subsection{A Dirichlet Process Mixture Model \label{S:Model}}
The above is formalised using a latent DPMM to model $f(\theta)$ \citep[see][for further details]{Neal2000, Walker2007}. To facilitate interpretation and sampling, we introduce the latent allocation variable $c_i$ denoting the cluster within the infinite Dirichlet process (DP) mixture to which the $i^{\textrm{th}}$ object belongs. Also, let $\boldsymbol{\zeta} = (\zeta_1, \zeta_2, \ldots)$, with $\zeta_j = (\phi_j, \tau_j)$ the mean and precision of the $j^{\textrm{th}}$ cluster within the DPMM. Our model then becomes:
\begin{align*}
&X_i | \theta_i \sim N(\mu(\theta_i), \sigma_i^2) \;\;\mathrm{for}\, i = 1, \ldots, n, \\
&\theta_i | c_i, \boldsymbol{\zeta} \sim N(\phi_{c_i}, \tau_{c_i}^{-1}) \;\;\mathrm{for}\, i = 1, \ldots, n \;\;\mathrm{and} \\
&\zeta_j \sim G_0 \;\;\mathrm{for} \, j = 1, 2, \ldots,  \\
&P(c_i = j) = w_j \;\;\mathrm{for} \, j = 1, 2, \ldots \;\; \mathrm{and} \;\; i = 1, \ldots, n. 
\end{align*}
We place a stick breaking prior on the mixture weights, so $w_1 = v_1$ and $w_j = v_j \prod _{l<j}(1-v_l)$ for $j > 1$, with $v_j \sim \mathrm{Beta}(1,\alpha )$ for $j = 1, 2, \ldots$. In the case of the P\'olya urn updating of the DPMM \citep{Neal2000}, these mixing weights are integrated out; while in the slice sampling approach \citep{Walker2007} they remain as explicit variables describing the current state of the model. Finally, to enable conjugacy in our updating, $G_0$ is such that
\begin{align*}
(\phi_j, \tau_j) | \mu_{\phi} &\sim NormalGamma(\mu_{\phi}, \lambda, \nu_1, \nu_2).
\end{align*}

\paragraph{Hyperparameters and hyperpriors}
We also place hyperpriors on $\mu_\phi$ (the overall cluster centering) and the DP concentration parameter $\alpha$ (determining the number of clusters we expect to observe amongst our $n$ sampled objects):
\begin{align*}
\mu_{\phi} &\sim N(\xi, \psi^{-1}), \\
\alpha &\sim Gamma(\eta_1, \eta_2).
\end{align*}
The values of $\xi$, $\psi$, $\eta_1$ and $\eta_2$; along with those of $\lambda$, $\nu_1$ and $\nu_2$ used in our prior on the mean and precision of the calendar ages in each cluster are fixed, but set at levels that adapt to the initial \rC observations $X_1, \ldots, X_n$. See Section \ref{S:ParamChoices} for further details.

\subsubsection{Notation Definition}
\paragraph{Observed/known variables}
\begin{align*}
& \mathbf{X_n} = (X_1, \ldots, X_n) \textrm{ --- observed \rC determinations of objects,} && \\
& \sigma_1, \ldots, \sigma_n \textrm{ --- sd on \rC determinations,} &&\\
& m(\theta) \textrm{ --- mean of radiocarbon calibration curve at calendar age $\theta$,} && \\
& \rho(\theta) \textrm{ --- standard deviation of radiocarbon calibration curve at calendar age $\theta$.} &&
\end{align*}

\paragraph{Unobserved variables of immediate interest}
\begin{align*}
&\boldsymbol{\theta} = (\theta_1, \ldots, \theta_n) \textrm{ --- underlying calendar ages of objects}. &&
\end{align*}

\paragraph{Unobserved latent variables --- corresponding to $f(\theta)$}
\begin{align*}
&\mathbf{c} = (c_1, \ldots, c_n) \textrm{ --- cluster identifier for each object}, &&\\
&\boldsymbol{\phi} = (\phi_{1}, \ldots) \textrm{ --- mean of each normal cluster}, && \\
&\boldsymbol{\tau} = (\tau_{1}, \ldots) \textrm{ --- precision of each normal cluster}, && \\
&\mathbf{w} = (w_{1}, \ldots) \textrm{ --- mixing weights of each cluster (not used in P\'olya urn updating approach)}, && \\
&\alpha \textrm{ --- DP concentration prior parameter}, &&\\
&\mu_{\phi} \textrm{ --- overall centre of clusters}. &&
\end{align*}
The ingenuity of the two approaches to DPMM updating means that, while it may appear we have to sample the entire set of $\boldsymbol{\phi}$, $\boldsymbol{\tau}$ and $\mathbf{w}$, we are in fact required to store only a finite set of values at any step dependent upon the current number of clusters explicitly modelled in the DPMM.

\paragraph{Hyperparameters --- fixed}
\begin{align*}
& (\lambda, \nu_1, \nu_2) \textrm{ --- influence cluster means and precisions, $(\phi_j, \tau_j) | \mu_{\phi} \sim NormalGamma(\mu_{\phi}, \lambda, \nu_1, \nu_2)$}, && \\
& (\xi, \psi) \textrm{ --- mean and precision on overall centering, $\mu_\phi \sim N(\xi, \psi^{-1})$}, &&\\
& (\eta_1, \eta_2) \textrm{ --- shape and rate on prior for DP concentration, $\alpha \sim \Gamma(\eta_1, \eta_2)$.}
\end{align*}
These hyperparameters are set at levels informed by our observed \rC determinations $X_1, \ldots, X_n$, but are then considered fixed within the MCMC sampler.

\subsection{Gibbs Sampling \label{S:GibbsSamplerOverview}}
The current state of our sampler is specified by $(\boldsymbol{\theta}, \mathbf{c}, \boldsymbol{\phi}, \boldsymbol{\tau}, \mathbf{w}, \alpha, \mu_\phi)$\footnote{or without $\mathbf{w}$ in the case of the P\'olya Urn DPMM approach where the mixing weights are integrated out.}. Updating is performed within an overall Gibbs MCMC scheme by sampling in turn from:
\begin{flushleft}
\begin{tabular}{@{\hspace{10em}}l@{}ll}
\multicolumn{3}{l}{1. Update $ \theta_i | X_i, f(\theta)$:} \\
& \multicolumn{2}{l}{\hspace{5em} $\theta_i | X_i, c_i, \phi_{c_i} \tau_{c_i} \quad \textrm{for } i = 1, \ldots, n$}  \\
\multicolumn{3}{l}{2. Update DPMM $f(\theta) | \boldsymbol{\theta}, \alpha, \mu_\phi$:} \\
& Either P\'olya Urn & Or Slice Sampling \\
& $c_i | \theta_i, \boldsymbol{\phi},  \boldsymbol{\tau}, \mathbf{c}_{-i} \quad \textrm{for } i = 1, \ldots, n$ & $\mathbf{w} | \mathbf{c}, \alpha$ \\
& $(\boldsymbol{\phi}, \boldsymbol{\tau}) | \boldsymbol{\theta}, \mathbf{c}, \mu_\phi$ &  $c_i | \theta_i, \boldsymbol{\phi},  \boldsymbol{\tau}, \mathbf{w} \quad \textrm{for } i = 1, \ldots, n$ \\
&& $(\boldsymbol{\phi}, \boldsymbol{\tau}) | \boldsymbol{\theta}, \mathbf{c}, \mu_\phi$ \\
\multicolumn{3}{l}{3. Update DPMM hyperparameters:} \\
& \multicolumn{2}{l}{\hspace{5em} $\alpha | \mathbf{c}$} \\
& \multicolumn{2}{l}{\hspace{5em} $\mu_{\phi} | \boldsymbol{\phi}, \boldsymbol{\tau}$} \\
\end{tabular}
\end{flushleft}
Note how the introduction of the allocation variable $c_i$ in the DPMM greatly simplifies the conditional calibration of each object, i.e., $ \theta_i | X_i, f(\theta)$, in step 1. We do not need to consider the complete prior $\theta_i \sim f(\theta)$ since, conditional on the allocation $c_i$ of the object, we know which specific cluster/component of the infinite mixture it arises from, i.e., $\theta_i \sim N(\phi_{c_i}, \tau_{c_i}^{-1})$. All the conditionals in this Gibbs scheme, except $\alpha | \mathbf{c}$ which is updated by Metropolis-Hastings, are directly sampled. This makes the algorithm quick and hopefully improves mixing: 

\paragraph{Step 1: Updating $\theta_i | X_i, c_i, \phi_{c_i} \tau_{c_i}$}
Given $c_i$, $\theta_i$ belongs to the $c_i^\textrm{th}$ component of our infinite mixture, i.e., our prior reduces to $\theta_i | c_i \sim N(\phi_{c_i}, \tau_{c_i}^{-1})$. We therefore wish to sample from the posterior distribution with density
\begin{align*}
f(\theta_i | X_i, \phi_{c_i}, \tau_{c_i}) &\propto f(X_i | \theta_i) f(\theta_i | c_i, \phi_{c_i}, \tau_{c_i}) \\
&= \varphi(X_i; m(\theta_i), \rho(\theta_i)^2 + \sigma^2_i) \varphi(\theta_i; \phi_{c_i}, \tau_{c_i}^{-1}),
\end{align*}
where $\varphi(x; \mu, \sigma^2)$ is the pdf of a normal with mean $\mu$ and variance $\sigma^2$. Due to the non-monotonicity of the calibration curve mean $m(\theta)$, this conditional is a non-standard distribution and likely multimodal, meaning that a Metropolis-Hastings step may introduce mixing difficulties. We aim to overcome this using a slice sampling approach \citep{Neal2003}. To update $\theta_i$, we proceed via a three step procedure:
\begin{enumerate}
\item Sample an auxiliary variable $z = \log p(\theta_i) - e$ where $e \sim Exp(1)$. This defines a horizontal slice $S = \{\theta : z < \log(p(\theta)) \}$.
\item Find an interval $I = [L,R]$ around $\theta_i$ that contains all, or most of this slice $S$
\item Draw a new $\theta_i^\star$ from the part of the slice in this interval
\end{enumerate}
There are multiple ways to find, and sample, from the interval $I$. Our implementation uses the ``stepping-out and shrinkage'' procedure \citep[see][for details]{Neal2003}. Here an initial interval of width $w$ is randomly positioned around $\theta_i$, and then expanded in steps of size $w$ until both ends are outside the slice. Then $\theta_i^\star$ is found by picking uniformly from the ``stepped-out'' interval until a point in the slice is found. Points picked outside the slice are used to ``shrink'' the interval.

\paragraph{Step 2: Updating DPMM - Either by P\'olya Urn or Slice Sampling}
As explained in Section \ref{S:GibbsSamplerOverview}, we consider two different schemes to update the latent DPMM --- a P\'olya Urn approach \citep{Neal2000} which integrates out the mixing weights $\mathbf{w}$; and a slice sampling approach in which they are explicitly retained \citep{Walker2007}. The specific details can be found in \citet{Neal2000} and \citet{Walker2007}.  

\paragraph{Step 3a: Updating $\alpha | \mathbf{c}$}
This update is performed using Metropolis-Hastings. Given the allocations $\mathbf{c}$, the Chinese restaurant process analogy for a DP, and its exchangeability, provide a likelihood,
$$
L(\alpha; \mathbf{c}) \propto \frac{\alpha^{n_c} \prod_{j=1}^{n_c} (n_{c,j} -1)!}{\frac{(\alpha+n-1)!}{(\alpha-1)!}},
$$
where $n_c$ is the total number of clusters and $n_{c,j}$ is the number of elements in cluster $j$. We therefore add an update step:
\begin{itemize}
 \item Sample $\alpha^\star \sim N^+(\alpha, \sigma_\textrm{prop}^2)$ where $N^+()$ is a truncated normal restricted to $(0, \infty)$.
 \item Accept with probability (adjusted due to non-symmetric proposal)
 $$
 \min \left\{ 1, \frac{\pi(\alpha^{\star})}{\pi(\alpha)} \frac{\Phi(\alpha/\sigma_\textrm{prop})}{\Phi(\alpha^\star/\sigma_\textrm{prop})} \frac{L(\alpha^\star; \mathbf{c})}{L(\alpha; \mathbf{c})} \right\}
 $$
 where $\pi(\cdot)$ is the density of the Gamma prior and $\Phi$ the cdf of a standard normal.
\end{itemize}

\paragraph{Step 3b: Updating $\mu_{\phi} | \boldsymbol{\phi},  \boldsymbol{\tau}$}
We have $(\phi_c | \mu_{\phi}, \boldsymbol{\tau}) \sim N(\mu_{\phi}, 1 / (\lambda \tau_c))$ and so, with our conjugate prior $\mu_\phi \sim N(\xi, \psi^{-1})$, we can sample directly from:
$$
\mu_{\phi} | \boldsymbol{\phi}, \boldsymbol{\tau}^2 \sim N\left( \frac{\xi \psi +  \sum_c \tau_c \phi_{c}}{\psi + \sum_c \tau_c}, \frac{1}{\psi + \sum_c \tau_c} \right)
$$

\subsection{Outputs}
Our sampler provides three outputs of particular interest: 
\paragraph{Calendar Ages} We obtain, for each object, the posterior distributions of its calendar age, i.e., $\theta_i | X_1, \ldots, X_n$. These estimates use the joint information provided by all the \rC determinations (as opposed to solely the \rC determination, $X_i$, of the single object) on the understanding the calendar ages of the objects are related. Incorporating this joint information, done without the need to specify a parametric phase model,  should improve calendar age estimation.     

\paragraph{Density Estimate to Summarise Objects  \label{S:Predictive}}
We also obtain the predictive distribution for the calendar age of a new, as yet undiscovered, object, i.e.,  $\theta_{n+1} | X_1, \ldots, X_n$. This density estimate summarises the calendar ages of all $n$ objects. It is generated using the posterior sampled values $(\mathbf{c}^\star, \boldsymbol{\phi}^\star, \boldsymbol{\tau}^\star, \mathbf{w}^\star, \alpha^\star, \mu_\phi^\star)$ of the DPMM component of our MCMC sampler. Let $f(\theta_{n+1} | \mathbf{X_n})$ denote the predictive density for the calendar age $\theta_{n+1}$ conditioned on the observed \rC determinations $\mathbf{X_n} = (X_1, \ldots, X_n)$, and $\mathcal{G}| \mathbf{X_n}$ be our posterior for the DP given $\mathbf{X_n}$. Then, for a probability measure $G$ drawn from $\mathcal{G}| \mathbf{X_n}$
\begin{align*}
f(\theta_{n+1} | \mathbf{X_n}) &= \int \varphi(\theta_{n+1} | \zeta_{n+1}) \pi(d\zeta_{n+1} | \mathbf{X_n}) \\
&= \int \int \varphi(\theta_{n+1} | \zeta_{n+1}) G(d\zeta_{n+1}) \mathcal{G}(dG | \mathbf{X_n}).
\end{align*}
Given a set of sampled values from the posterior $(\mathbf{c}^\star, \boldsymbol{\phi}^\star, \boldsymbol{\tau}^\star, \mathbf{w}^\star, \alpha^\star, \mu_\phi^\star)$, then $\zeta_{n+1}$ is drawn from either:
\begin{itemize}
\item One of the current clusters modelled in the DPMM --- with probabilities according to the current truncated, finite-length, vector of explicitly calculated mixture weights $\mathbf{w}^\star$ (or, for the P\'olya Urn, dependent upon the current allocations $\mathbf{c}^\star$ using the Chinese restaurant process analogy). The means and precisions $(\boldsymbol{\phi}^\star, \boldsymbol{\tau}^\star)$ for these clusters are known. After integration, this contributes a finite mixture of normals (with known probabilities) to the predictive for $\theta_{n+1}$.
\item A currently unmodelled cluster in the DPMM --- with a probability dependent upon the sum of the truncated mixing weights, $1 - \sum \mathbf{w}^\star$ (or, for the P\'olya Urn, dependent upon the current allocations $\mathbf{c}^\star$). In such a case, we draw from the base distribution, i.e., $\zeta_{n+1} \sim NormalGamma(\mu^\star_{\phi}, \lambda, \nu_1, \nu_2)$. After integration, this contributes a scaled t-distribution to the predictive for $\theta_{n+1}$.
\end{itemize}
Hence, for an individual realisation of measure $G \sim \mathcal{G}| \mathbf{X_n}$ and specified by $(\mathbf{c}^\star, \boldsymbol{\phi}^\star, \boldsymbol{\tau}^\star, \mathbf{w}^\star, \alpha^\star, \mu_\phi^\star)$, the predictive  $\int \varphi(\theta_{n+1} | \zeta_{n+1}) G(d\zeta_{n+1})$ is a finite mixture of normals (parameterised by the current clusters in that realisation) and a single $t$-distribution corresponding to the possibility of a new cluster. The predictive density estimate $f(\theta_{n+1} | \mathbf{X_n})$ is found by computing this finite mixture averaged over multiple posterior samples $(\mathbf{c}^\star, \boldsymbol{\phi}^\star, \boldsymbol{\tau}^\star, \mathbf{w}^\star, \alpha^\star, \mu_\phi^\star)$.

\paragraph{Number of Clusters and Object Allocation}
Finally, we obtain the allocation $\mathbf{c}$ of the $n$ sampled objects to clusters within the DPMM. If we believe the underlying individual clusters in the DPMM to have inherent meaning in terms of representing genuine and distinct periods of site usage, as opposed to simply providing a tool to enable a non-parametric density estimate, this information may be archaeologically useful. In particular, the posterior for $\mathbf{c}$ allows an estimate of the potential number of, in our case normally distributed, phases observed. 

\subsection{Choice of parameters and hyperpriors \label{S:ParamChoices}}
Our hyperparameters play an important role in determining the location and spread of the individual clusters in the DP mixture used to construct our estimate of $f(\theta)$, as well as the level of clustering seen in the objects. Selection of these hyperparameters could be done by a user based upon their personal beliefs regarding the set of samples they are studying. However we see a benefit in suggesting default values. These are set out below and are based upon discussion with archaeological experts. The proposed values intend to reflect their experience, yet remain relatively uninformative and scale invariant. These defaults were the values used in our simulation study, and for all three of our real life examples in Section \ref{S:RealExamples}.

\subsubsection{Choice of prior on mean and precision of a given DP cluster}
We place a normal-gamma prior on the mean and precision of any individual calendar age cluster, i.e., $(\mu_{j}, \tau_{j}) \sim NG(\mu_{\phi}, \lambda, \nu_1, \nu_2)$. This requires specification of four parameters: $\mu_{\phi}$ denotes the global central tendency of all the data, i.e., the expected mean age of a cluster; $\lambda$ influences how far the centre of an individual cluster can lie from this global centre; and $\nu_1$ and $\nu_2$ affect the length/spread of any individual cluster. Any particular cluster will have calendar ages $\theta_i \sim N(\mu_{c_i}, \tau_{c_i}^{-1})$, hence if we can specify our prior beliefs about the lengths and locations of calendar age clusters in an archaeological setting this can be used to guide suitable parameter choices. 

To select our proposed default and scale invariant hyperparameter values, we first perform a fast and approximate calibration of each \rC determination $X_1, \ldots, X_n$ independently against a coarse sampling of the calibration curve. We calculate, on a rough calendar age grid, the likelihood $f_i(\theta_i | X_i) \propto \varphi(X_i ; m(\theta), \sigma_i^2 + \rho^2(\theta_i))$, where $\varphi(\cdot; \mu, \sigma^2)$ denotes the normal density with mean $\mu$ and variance $\sigma^2$. This allows us to approximately determine $\tilde{\theta}_i$, the most likely independent (MAP) calendar age for each object. From these $\tilde{\theta}_i$, we calculate the overall range of calendar ages, $\textrm{range}(\tilde{\theta})= (\max_i \tilde{\theta}_i  - \min_i \tilde{\theta}_i)$ which we will use to restrict the centres of each component our DPMM. We also calculate the maximum absolute deviation, $\textrm{mad}(\tilde{\theta})$, to obtain a robust estimate of the overall spread of calendar ages which we will use to restrict the maximum spread of any individual cluster.  

\paragraph{The spread of a cluster --- selecting $\nu_1$ and $\nu_2$}
The parameters $\nu_1$ and $\nu_2$ specify our prior on the spread,  $\sigma_{\textrm{cluster}, j}$, of any individual cluster. Within any cluster, the prior variance on the calendar ages $\theta$ drawn from that component is, 
$$
\sigma_{\textrm{cluster}, j}^2 = \tau_{j}^{-1}  \sim \Gamma^{-1}(\nu_1, \nu_2).
$$
We wish to permit a broad range of clusters with calendar age spreads that capture the potential for both short lived, intense periods of activity; and longer periods of more steady use. We select $\nu_1 = 0.25$ and $\nu_2 = \textrm{mad}(\tilde{\theta})^2 \times \nu_1 / 100$, corresponding to a very positively-skewed prior, with heavy right hand tails, on the variance of the calendar ages in any cluster. This gives a prior for the spread $\sigma_{\textrm{cluster}, j}$ of any individual cluster that has an upper 75\% percentile of approximately $\textrm{mad}(\tilde{\theta})$ cal yrs --- a sensible upper bound considering the extreme case where the samples arise from a single cluster. This prior still allows narrow cluster spreads, in the case where $\textrm{mad}(\tilde{\theta}) = 1000$ cal yrs, the lower 5\% quantile for the prior spread of a cluster is approximately 45 cal yrs.  

\paragraph{The location of a cluster --- selecting $\lambda$ and $\mu_\phi$}
Conditional on the value of $\tau_j$, our choice of conjugate Normal-Gamma prior results in the mean calendar age of an individual cluster,
\begin{align*}
\mu_j | \tau_j \sim N(\mu_{\phi}, \frac{1}{\lambda \tau_j}).
\end{align*}
Hence wider (lower precision) clusters can be located further from the central $\mu_{\phi}$ than narrower (higher precision) clusters. Here, $\lambda$ influences how far an individual cluster can be located from the central $\mu_{\phi}$, the smaller its value the further away that clusters may lie. We select an uninformative prior by setting $\lambda = (100/\textrm{range}(\tilde{\theta}))^2$, so that a cluster with spread $\sigma_{\textrm{cluster}, j} = 50$ cal yrs should, according to our prior with 95\% probability, lie within approximately $\textrm{range}(\tilde{\theta})$ calendar years either side of the central $\mu_{\phi}$. We also place a hierarchical prior on $\mu_{\phi}$,
\begin{align*}
\mu_{\phi} \sim N(\xi, \psi^{-1}).
\end{align*}
We set $\xi$ to be the median of the preliminary $\tilde{\theta}_i$ calendar age estimates; and, to provide a conservative estimate of the overall range of the calendar ages, we set $\psi^{-1} = \textrm{range}(\tilde{\theta})^2$. 

Note that, if a user believes that the spread of a cluster should not be related to its distance from the central $\mu_\phi$ they might, instead of our joint Normal-Gamma prior, wish to place independent Normal and Gamma priors on the mean $\phi_j$ and precision $\tau_j$ of a cluster respectively. This alternative should not substantially affect implementation, simply requiring the cluster means and precisions to be updated in separate and independent steps, as well as the appropriate modification to step 3b updating $\mu_{\phi} | \boldsymbol{\phi}$.  

\subsubsection{Choice of prior on $\alpha$}
The parameter $\alpha$ used to determine the mixing weights in our DPMM affects the clustering of the objects. A small value of $\alpha$ means that the DP is concentrated and the method will have a preference to group the objects into fewer, but larger, distinct clusters; conversely a large value of $\alpha$ means the Dirichlet process is less concentrated resulting in a greater number of separate clusters. Given a particular value of $\alpha$ and a number of observations $n$, the expected number of distinct clusters $k_n = \sum_{i=1}^n \alpha / (\alpha + i - 1)$. If archaeological prior information on the number of clusters in our set of \rC determinations exists, this relationship could be used to select a suitable value for $\alpha$. 

\begin{figure}
  \centering
      \includegraphics[width=0.99\textwidth, keepaspectratio]{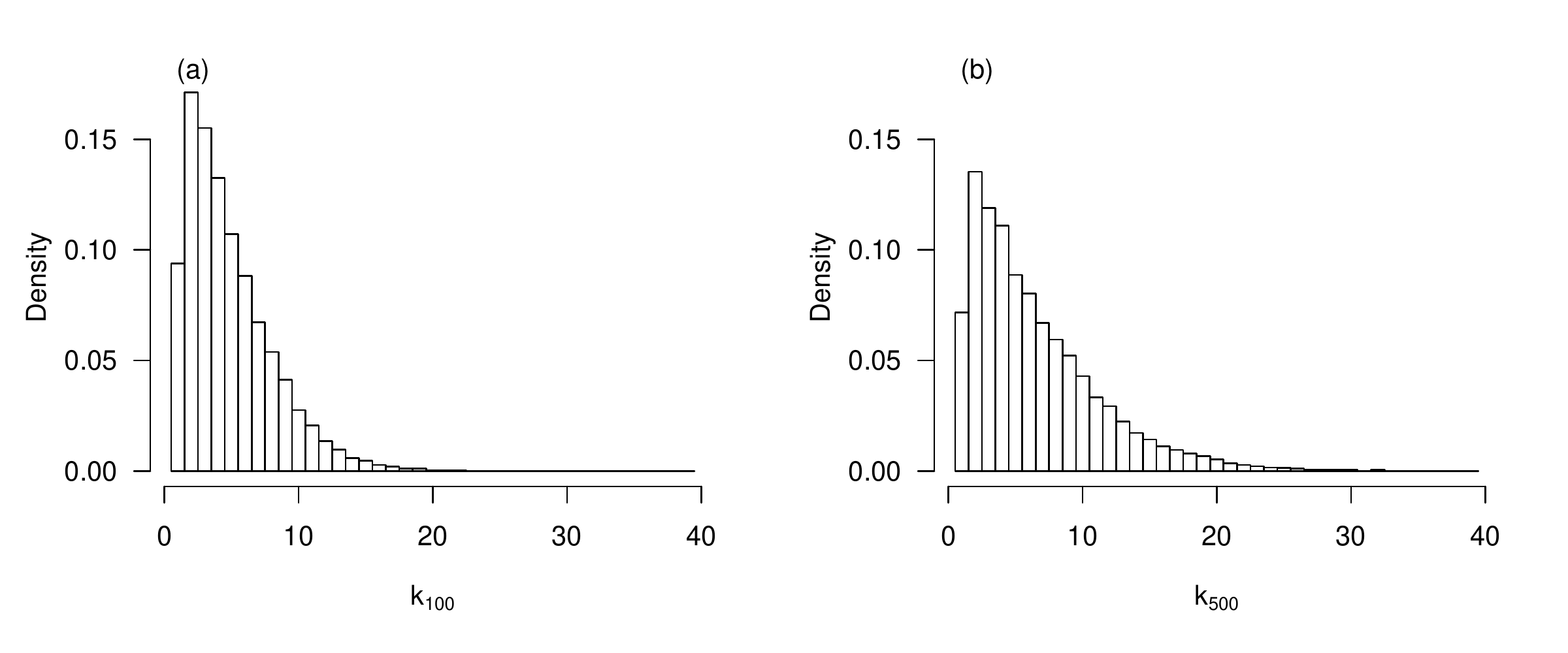}
  \caption{\textit{Induced marginal prior on $k_{100}$ and $k_{500}$, the expected number of distinct clusters observed in $100$ and $500$ sampled objects respectively, with our suggested hyperprior $\alpha \sim \Gamma(1, 1)$.} \label{F:ExpectedClusterNum}}
\end{figure} 

In the absence of specific information on the number of clusters, and to allow greater adaptation to the data under consideration, we suggest placing a hyperprior on $\alpha \sim \Gamma(\eta_1, \eta_2)$. Our examples set $\eta_1 = 1$ and $\eta_2 = 1$ to provide a relatively uninformative prior. In Figure \ref{F:ExpectedClusterNum}, we show the marginal prior this induces on $k_{100}$ and $k_{500}$ --- the expected number of distinct clusters for 100 and 500 \rC determinations respectively. This hyperprior provides an approximate 95\% prior probability that $k_{100} \in [1,13]$ and $k_{500} \in [1,19]$. We also investigated use of a log-normal hyperprior for $\alpha$ but, on some of our examples, its very heavy tails did not penalise extreme $\alpha$ values sufficiently leading to implausible numbers of clusters.

\section{Simulation Study \label{S:SimStudy}}
\subsection{Improving Calibration}
We performed a simulation study to investigate the improvement in calendar age estimation possible with our non-parametric joint \rC calibration approach, compared with independent calibration of each \rC determination where no information is shared between samples. We considered three families for the underlying calendar age distribution $f(\theta)$ ---  a single normal distribution/phase, a mixture of three underlying normal distributions/phases, and a uniform phase. Given our chosen $f(\theta)$, we sampled calendar ages $\theta_i$, for $i = 1, \ldots, n$; and corresponding \rC determinations $X_i \sim N(\mu(\theta_i),  \sigma_\textrm{obs}^2)$ using the IntCal20 calibration curve \citep{Reimer2020} and a typical laboratory uncertainty $\sigma_\textrm{obs} = 25$ cal yrs. We then aimed to estimate each $\theta_i$ given the set of radiocarbon determinations $\mathbf{x} = (x_1, \ldots, x_n)$ and the chosen uncertainty $\sigma_\textrm{obs}$.

We tested three calibration approaches: our proposed non-parametric Bayes method, using both the P\'olya urn and slice sampling DPMM updates, where joint information is shared between the related objects to obtain calendar age estimates $\hat{\theta}_i | x_1, \ldots, x_n$; and calibration of each \rC determination $x_i$ independently of the others using an uninformative prior on its calendar age and where no information is shared between the samples, i.e., $\hat{\theta}_i | x_i$. This latter method is the approach of most current \rC users when they do not wish to select a specific parametric phase model. For our simulation study, our MCMC sampler was run for 10,000 iterations, with the first 5,000 iterations discarded as burn-in, before being thinned to every $5^{\textrm{th}}$ iteration. This relatively small number of MCMC iterations is only used for this simulation study on calibration losses where computational speed is needed. For real-life use of the method for either calibration or density estimation, we recommend a longer MCMC run and a greater number of posterior samples to ensure stability in our estimates. All our later individual work is based on MCMC run of 50,000 iterations and posterior samples of size 5000. As recommended by \citet{Hastie2015} we have initiated our sampler with a greater number of clusters than in the underlying $f(\theta)$ --- in our studies we initiated the DP samplers with 10 clusters. 

\begin{figure}[t!]
  \centering
      \includegraphics[width=0.99\textwidth, keepaspectratio]{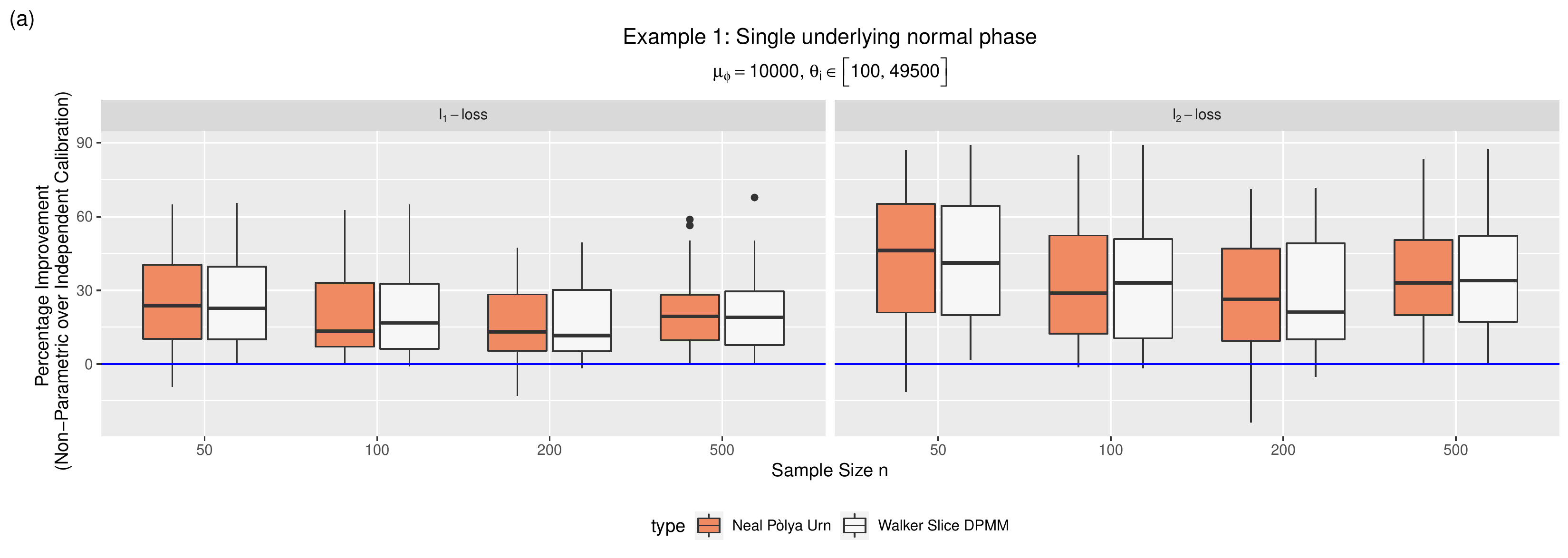}
      \includegraphics[width=0.99\textwidth, keepaspectratio]{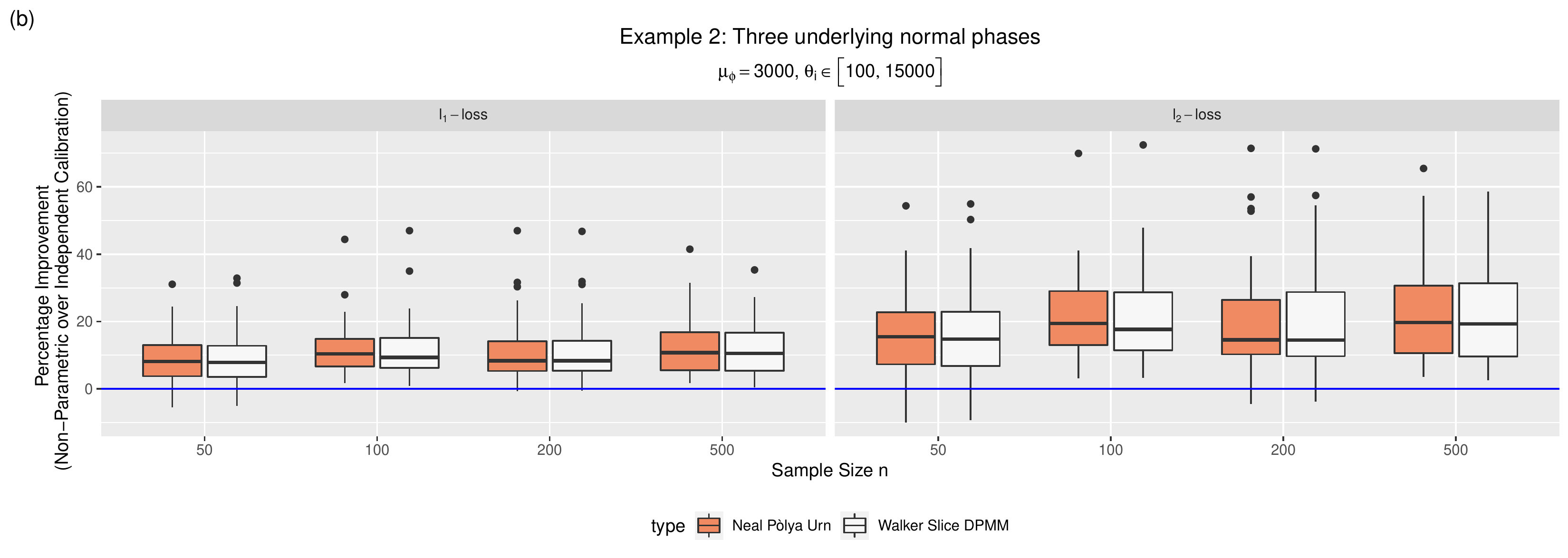}
      \includegraphics[width=0.99\textwidth, keepaspectratio]{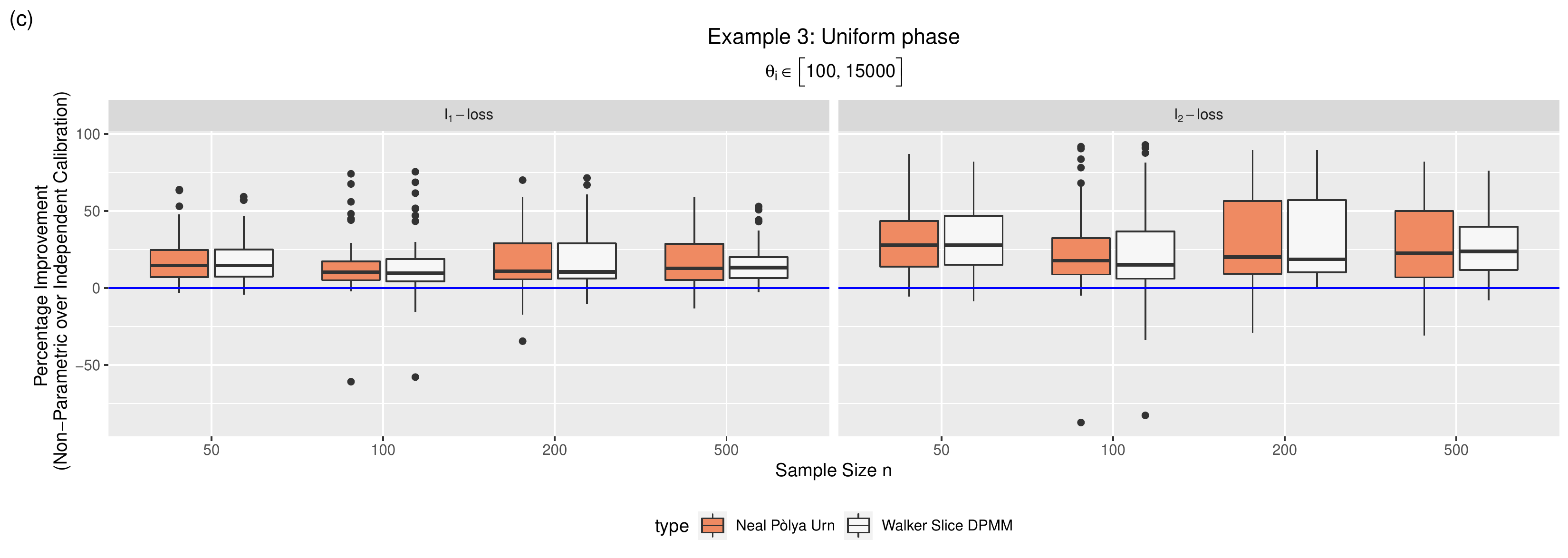}
      \caption{\textit{Box-plot of percentage improvements in calendar age estimation of $\theta_i$ achieved when using our joint non-parametric approach, compared to independent calibration of each \rC determination, if the calendar ages of the samples are known to be related. The horizontal blue line corresponds to zero improvement. Runs lying above this indicate the DPMM approach reduced the overall loss, and hence improved calibration.} \label{F:SimStudy}}  
\end{figure}

We investigated the performance for $n = 50, 100, 200$ and $500$ \rC determinations. For a given $n$, we performed 50 runs, each sampling a new and different $f(\theta)$ from the chosen family. Each approach to calibration provided a posterior distribution $\hat{\theta}_i | \mathbf{x}$ for the calendar age of the $i^{th}$ determination. We quantified the quality of calibration for a particular method using the sample average posterior expected losses:
\begin{align*}
\textrm{Absolute ($l_1$) loss:}& \quad  \frac{1}{n} \sum_{i=1}^n \EX \big| [ \hat{\theta}_i | \mathbf{x} ] - \theta_i \big|; \\
\textrm{Mean-squared ($l_2$) loss:}& \quad \frac{1}{n} \sum_{i=1}^n \EX \big( [ \hat{\theta}_i | \mathbf{x} ] - \theta_i \big)^2.
\end{align*}
To assess the potential improvement in calibration accuracy offered by our non-parametric joint approach we compared, for each simulation run, the posterior losses of these methods against those obtained when calibrating each object independently:
$$
\textrm{Non-Parametric Improvement} = 100 \times \left( 1 - \frac{\textrm{Loss with Non-Parametric Bayes}}{\textrm{Loss if calibrate independently}} \right) \%.
$$
We present, in Figure \ref{F:SimStudy}, box-plots of the improvements in the joint calendar age estimation loss for each of our 50 simulation runs obtained using our non-parametric approach compared with independent calibration of each \rC sample. In Table \ref{T:SimStudy} we also provide the percentage of times (out of the 50 runs) where our non-parametric approaches offered an improvement over independent calibration; the mean improvement they offered over the 50 runs; and the maximum and minimum improvement.

\subsubsection{Underlying distributions $f(\theta)$}
\paragraph{Calendar age distribution 1: Single normal phase}
For each $n = 50, 100, 200, 500$, and each run $k = 1, \ldots, 50$, we sample underlying calendar ages from a single normal phase and their corresponding \rC determinations:
\begin{align*}
& \tau \sim Gamma(1, 10000), \;\; \phi \sim N(10000, \frac{100}{\tau^2}); \\
&\theta_i \sim N(\phi, \frac{1}{\tau}) \;\; \textrm{and }
X_i \sim N(m(\theta_i), \sigma_\textrm{obs}^2 + \rho(\theta_i)^2) \;\; \textrm{for } i = 1, \ldots, n.
\end{align*}
We set $\sigma_\textrm{obs}$, the laboratory measurement uncertainty to be 25 $^{14}C$ years, a typical level of  accuracy a laboratory might provide; while $m(\theta)$ and $\rho(\theta)$ are the published posterior mean and standard deviation on the IntCal20 calibration curve at calendar age $\theta$.  

\paragraph{Calendar age distribution 2: Mixture of three normal phases during last 15,000 cal years}
For each $n$, and run $k$, we sample underlying calendar ages from three normal phases and their corresponding \rC determinations:
\begin{align*}
&\tau_j \sim Gamma(1, 10000) \textrm{ and } \phi_j \sim N(3000, \frac{100}{\tau_j^2}) \; \textrm{for } j = 1, 2, 3; \;\; w_1, \ldots, w_3 \sim Dir(1,1,1); \\
&\theta_i \sim \sum_{j=1}^3 w_j \varphi(\theta; \phi_j, \frac{1}{\tau_j}) \;\; \textrm{and }
X_i \sim N(m(\theta_i), \sigma_\textrm{obs}^2 + \rho(\theta_i)^2) \;\; \textrm{for } i = 1, \ldots, n.
\end{align*}
Here, we also restrict the calendar ages $\theta_i$ to lie within the last 15,000 cal yrs. In this time period the calibration curve is based on densely-sampled and high-precision tree-ring \rC determinations. The additional detail these high-precision, frequently annually-sampled, measurements provide on past \rC levels results in a highly non-monotonic calibration curve (see Figure \ref{F:MixtureDensityReconNormal}) which should particularly test our calendar age estimation. This more recent period, which includes the Holocene, is also the most highly interrogated by the radiocarbon community, and where they desire utmost precision in \rC calibration. If the above sampling creates any $\theta_i >$ 15,000 cal yrs, the entire sample is rejected and cluster creation restarted.

\paragraph{Calendar age distribution 3: Single uniform phase during last 15,000 cal years}
For each $n$, and each run $k$, we sample underlying calendar ages from a single uniform phase and their corresponding \rC determinations:
\begin{align*}
& S \sim U[100,14000], \;\; R \sim U[50, 1000]; \\
&\theta_i \sim U[S, S+R] \;\; \textrm{and }
X_i \sim N(m(\theta_i), \sigma_\textrm{obs}^2 + \rho(\theta_i)^2) \;\; \textrm{for } i = 1, \ldots, n.
\end{align*}
This also restricts the calendar ages $\theta_i$ to lie within the last 15,000 calendar years where the calibration curve is most non-monotonic and needs for \rC dating are typically greatest. The use of a uniform distribution should, at least when estimating the summarised predictive calendar age density, provide the most challenging test of our approach as we model our DP as mixtures of, light-tailed, normal distributions. It is particularly difficult to accurately approximate a uniform distribution with a mixture of normals.  

\subsubsection{Improvements in Calendar Age Estimation \label{S:SimStudyResults}}
The results of our simulation study are shown in Figure \ref{F:SimStudy} and Table \ref{T:SimStudy}. Joint calibration to estimate $\theta_i | x_1, \ldots, x_n$, using the entire set of \rC determinations, almost universally improves the accuracy of calendar age estimation when compared with independent calibration $\theta_i | x_i$. Across all three families of underlying calendar age distributions, over 95\% of the simulation runs show improved calendar age estimation using our proposed non-parametric approach that incorporates the information that the \rC determinations arise from a shared, but unknown, calendar age distribution. This rises to nearer 100\% when the underlying calendar age distribution is a mixture of normals (hence matching the components in the latent DPMM). Even when the joint approach does not offer an improvement in calendar age estimation, the loss is only very slightly worse than independent calibration. The benefit of joint calibration does not seem to be particularly affected by the number of \rC determinations. Even with just 50 \rC determinations, calendar age estimation is very significantly improved by joint calibration.    

For our single normal phase, our proposed joint calibration methods showed an improvement in calendar age estimation for almost all simulation study runs. The mean improvement in absolute error over independent calibration is in the order of 20\% for all $n$, while the maximum improvement offered is between 50 -- 60 \%. The mean improvement in squared error is between 30 -- 40 \%, with a maximum of c.a. 90\%. While this is perhaps the simplest test of our methods, since the underlying calendar age distributions are of a form which is easily approximated by our DPMM, they still need to recognise that the samples all arise from a single cluster. 

In the case of our mixture of three normals, we again see that joint calibration almost universally offers an improvement in the calendar age estimation. The mean improvement in calendar age estimation by jointly estimating $\theta_i | x_1, \ldots, x_n$ is less than in the case of a single normal phase but still highly significant compared with independent calibration $\theta_i | x_i$. Such a reduction is to be expected since, for this mixture family, the non-parametric methods need to recognise both that the samples arise from three distinct clusters and also estimate the mean and variance of each such cluster. Even if the DPMM assigns all the samples to their clusters correctly, there remain fewer \rC determinations in each cluster from which to precisely estimate its mean and variance. For both the P\'olya Urn and slice sampling approach to DPMM updating, the mean improvement in absolute error using non-parametric Bayes is in the order of 10\%; and the mean improvement in squared error is c.a. 20\%. 

Interestingly, when the underlying calendar age is uniform, joint calibration still offers large benefits in calendar age estimation. Non-parametric Bayes offered improvements in over 90\% of our simulation runs, with a mean reduction in absolute error of between 15 -- 20\%, and squared error of 25 -- 35\%, compared with independent calibration of the \rC determinations. These improvements are seen despite the uniform distribution being very difficult to estimate with a mixture of normals as chosen for the components in the latent DPMM. Our normal-distribution based DPMM is not particularly successful at estimating $f(\theta)$ when this underlying calendar age density is uniform, see Appendix \ref{S:AppendixUnifRecon}. When it comes to calibration however, even with a normal clusters in our DPMM, the method will still tend to shrink the individual calendar ages towards their mean. This typically provides improved age estimation. 

For the uniform, there some individual runs where non-parametric Bayes performed poorly --- after investigation, these were found to be runs where the calibration curve is flat for a prolonged period of time, and that the underlying calendar age density $f(\theta)$ covered only a short portion of this longer flat period. All the calibration methods perform poorly in such an instance as, given only the \rC determinations $X_1, \ldots, X_n$, identifiability of the underlying $f(\theta)$ and the individual $\theta_i$ is low due to the flatness of the calibration curve. However the non-parametric methods are particularly penalised since they shrunk the $\theta_i$ towards the middle of the flat period even though the true $f(\theta)$ lay towards one end. Joint calibration will always have this inherent danger. Fortunately, due to the underlying geoscience, there are not too many prolonged time periods where the calibration curve remains flat; and this risk can easily be identified by a user through comparison of the \rC determinations against the calibration curve as enabled by plots of the type in Figures \ref{F:MixtureDensityReconNormal} to \ref{F:BuchananDensity}.   
   
\subsection{Estimating the underlying calendar age density $f(\theta)$ \label{S:DensityRecon}}

As well as calibration, the DPMM methods simultaneously provide a prediction for the calendar age of a future object based upon the observed \rC determinations. Many users will wish to use this $\hat{f}(\theta)$ as a proxy for the population density/activity. Whether the DPMM's $\hat{f}(\theta)$ is able to identify the specific calendar age density behind the \rC samples will require careful consideration by any user. The DPMM summarisation method only has access to these observed \rC determinations and, intuitively, aims to provide an estimate $\hat{f}(\theta)$ of a calendar age distribution which could have generated these $X_1, \ldots, X_n$. Since the calibration curve is non-monotonic, there may be multiple calendar age densities which could generate the same $X_1, \ldots, X_n$. In such instances, the DPMM estimates of $f(\theta)$ would ideally cover the range of potential calendar age densities. Amongst these multiple possibilities, the true calendar age density may be unidentifiable. The predictive calendar age estimates should not therefore be considered a black-box.

Users will need to consider the nature of the underlying calibration curve in the period of interest if they wish to use $\hat{f}(\theta)$ as a proxy for population density/activity. In many instances, such inference may be reasonable but, in others, more care is required in interpretation of $\hat{f}(\theta)$. The probability intervals on $\hat{f}(\theta)$ provided by the DPMM method, which are not available with an SPD, should offer further guidance. Users should also note that, with a DPMM model that uses normally-distributed clusters, the estimator $\hat{f}(\theta)$ will likely work better for densities which can be well approximated by such a mixture of normals --- for example repeated colonisation (as perhaps caused by past climate variation) or the rise and fall of a civilisation/culture.

We provide examples of the estimates $\hat{f}(\theta)$ obtained by our DPMM, and SPDs, for an artificial calendar age density and a real-life example. Our MCMC samplers were run for 50,000 iterations with the first half discarded as burn-in. The samples were then thinned to every $5^{\textrm{th}}$ iteration so that the predictive estimates $\hat{f}(\theta)$ and their probability intervals were based on a final sample of size 5000. As with all our examples, we initiated our samplers with 10 initial clusters.

\subsubsection{Artificial Example --- Mixture of Normal Phases}

\begin{figure}[t]
  \centering
      \includegraphics[width=0.99\textwidth, keepaspectratio]{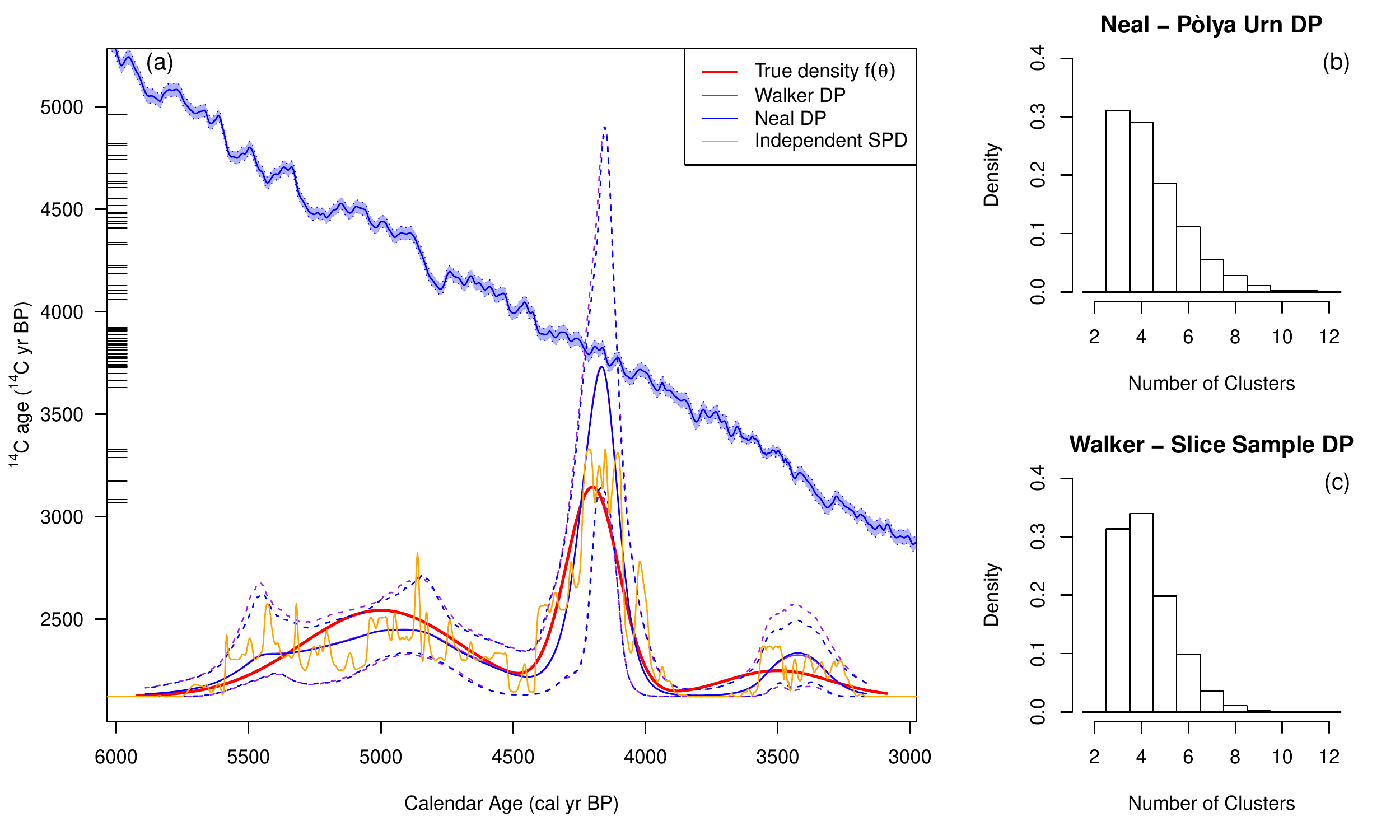}
  \caption{\textit{Estimates of $\hat{f}(\theta)$, based upon summarising a set of 100 \rC observations, when underlying samples have calendar ages drawn from a mixture of three normals: $\theta_i \sim f(\theta) = 0.1 N(3500, 200^2) + 0.4 N(4200, 100^2) + 0.5 N(5000, 300^2)$ cal yrs, for $i = 1, \ldots, 100$; and corresponding \rC determinations $X_i \sim N(m(\theta_i), \sigma_\textrm{obs}^2 + \rho(\theta_i)^2)$ \rCyrs. Panel (a) shows the 100 sampled \rC determinations as a ticked rug along the y-axis and the IntCal20 radiocarbon calibration curve (with shaded 95\% probability intervals) needed to convert them to calendar ages. Along the x-axis we plot the true underlying calendar density (red solid line) and the estimates $\hat{f}(\theta)$ obtained by calibrating and summarising the \rC determinations. The non-parametric Bayes estimate with P\'olya Urn DPMM updating is shown in blue (solid line, predictive mean; dashed line, 95\% predictive interval); and the slice sampling DPMM updating version in purple. The SPD estimate is plotted in orange. Panels (b) and (c) show the number of clusters in the DP mixture used to model the 100 \rC determinations in the P\'olya Urn and slice sampling versions respectively.} \label{F:MixtureDensityReconNormal}} 
\end{figure}

Figure \ref{F:MixtureDensityReconNormal} shows an illustrative DPMM estimate for the summarised calendar age density $\hat{f}(\theta)$ based upon 100 \rC determinations for which the underlying calendar ages were drawn from a mixture of three normals. We compare the true underlying calendar age density (shown in red) against the estimates obtained via our non-parametric Bayes, blue (P\'olya Urn DP updating) and purple (slice sampling); and the standard SPD estimate used in the \rC community. This SPD is created by initial calibration of each sample independently, $\theta_i | x_i$, without sharing any information across the \rC determinations, and then simply summing the individual and independent posterior calendar age densities, see Section \ref{S:SPDExplain} for details. The pointwise mean for the predictive density is shown with a solid line; 95\% pointwise credible intervals, based upon the individual realisations, are shown with dashed lines.   

For this mixture of normals, the non-parametric Bayesian DPMM summarisation approaches accurately reconstruct the underlying calendar age density $f(\theta)$. Both approaches to DPMM  updating identify the three distinct peaks and have credible intervals which encompass the true mixture density. The reconstruction of the central peak is somewhat sharper, and appears to be more heavily weighted, than the true density but still predominantly covers it within the credible intervals. We also see, in panels (b) and (c), that the non-parametric Bayes approaches have posteriors which place most probability on the \rC observations arising from three or four distinct calendar age clusters. The SPD estimate on the other hand is highly variable and fluctuates rapidly, due to the fine-scale wiggles and features of the calibration curve. It is challenging to identify the three distinct peaks in the underlying true density from this SPD estimate.      

\subsubsection{Real Example --- Population Change in the Basin of Mexico}
To illustrate the care required in interpreting the predictive density $\hat{f}(\theta)$ we also consider an example of \citet{Contreras2014} on the impact of European diseases on the indigenous population of the Basin of Mexico from c.a. AD 1000 -- 1900 \citep{McCaa2000}. From around AD 1428, this population grew rapidly as a result of the triple alliance of city states under the Aztec Empire and consequent regional population aggregation. The population peaked at around 1.2 million in AD 1520 just before the arrival of the Spanish conquistadors. These Europeans brought with them many diseases which decimated the indigenous population causing it to drop to around 170,000 by AD 1640, from which it began to slowly grow again.

\begin{figure}
  \centering
      \includegraphics[width=0.99\textwidth, keepaspectratio]{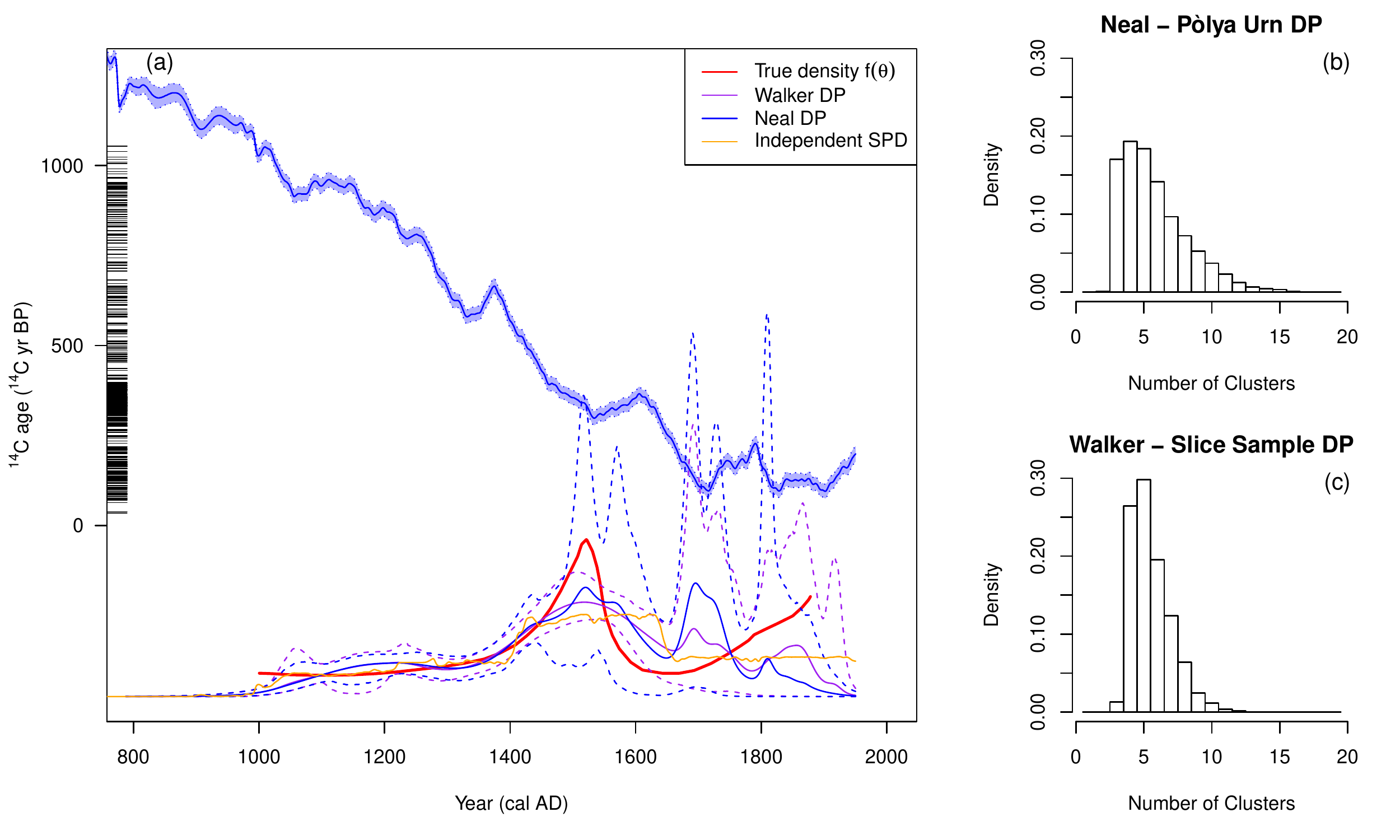}
  \caption{\textit{Estimates of $\hat{f}(\theta)$, based upon summarising 500 \rC observations, when the underlying samples have calendar ages $\theta_i$ that are drawn from a distribution proportional to the estimated population of the Basin of Mexico \citep{McCaa2000}. For each sample, the uncertainty in the  \rC measurement $\sigma_\textrm{obs} \sim \textrm{U}[20,40]$ \rC yrs, and $X_i \sim N(m(\theta_i), \sigma_\textrm{obs}^2 + \rho(\theta_i)^2)$. Individual panels as for Figure \ref{F:MixtureDensityReconNormal}.} \label{F:MixtureDensityReconMexico}}
\end{figure}

To test the ability of \rC summarisation to reconstruct demographic changes, \citet{Contreras2014} sampled calendar dates $\theta_1, \dots, \theta_n$  from AD 1000 -- 1900 with a density proportional to the population in that year. Figure \ref{F:MixtureDensityReconMexico} shows this implied calendar age density in red. The peak at AD 1520 corresponds to the maximum population of 1.2 million. Given each $\theta_i$, they sampled a \rC determination $X_i$ using the IntCal calibration curve. They then aimed to test whether summarising the resultant \rC determinations $X_1, \ldots, X_n$ returned the known population density. For our recreation, we sampled 500 calendar ages, and \rC determinations, from the underlying population-based density to reduce the effect of sampling variation on the estimates $\hat{f}(\theta)$.

This case study provides a somewhat pathological example for \rC summarisation but is helpful to show the care needed in interpretation and the benefits that the probability intervals on our DPMM estimate of $\hat{f}(\theta)$ provide compared to SPDs. As can be seen in Figure \ref{F:MixtureDensityReconMexico}, it is unfortunately not possible to recreate the underlying population density of the Basin of Mexico based upon our \rC observations. However, this is due to the nature of the calibration curve in this period rather than a failing of the DPMM method. Importantly the intervals on the DPMM $\hat{f}(\theta)$ help to flag this non-identifiability. The SPD estimate provides no such means to assess identifiability while still failing to reconstruct the underlying population changes.

There are a range of reasons for this failure to reproduce the Basin of Mexico demography. Firstly, the later rise in the population (from AD 1650) contains the Suess effect \citep{Suess1955}, the increase in atmospheric \nC due to the burning of fossil fuels since the industrial revolution around AD 1750. This effect results in a flat calibration curve from AD 1750 -- 1950 meaning that the calendar ages of \rC determinations from this period are fairly unidentifiable. Secondly, the example fails to consider edge effects. \citet{Contreras2014} simply cut off calendar ages outside a certain time range (including when the population is increasing significantly around AD 1880). Our normal clusters do not permit for such a hard cut-off and so estimation at these age boundaries will likely not be accurate. This is compounded because for \rC we cannot have calendar ages beyond AD 1950. Both these effects can be seen in our DPMM estimate. The \rC determinations from the later population rise could be equally likely to arise any time from AD 1800 onwards, or during the dip in the calibration curve around AD 1700. This unidentifiability is shown by the wide intervals on the DPMM predictive $\hat{f}(\theta)$ in this time period, and the peak around AD 1700. Thirdly, the main peak in the Mexican population leading up to AD 1520 is extremely narrow (\textit{c.a.}\ 50 cal yrs) and coincides with an inversion in the calibration curve. There are again two potential time periods which would create exactly the same set of \rC ages, AD 1520 and AD 1580. Again these two possibilities are shown by the intervals on the DPMM $\hat{f}(\theta)$. Any user seeing such wide, and variable, intervals should proceed with extreme caution in using the pointwise estimate for $\hat{f}(\theta)$ as a proxy for population/activity. 

 \subsubsection{A Modified Real Example --- Shifted Population Change in the Basin of Mexico}
\begin{figure}
  \centering
      \includegraphics[width=0.99\textwidth, keepaspectratio]{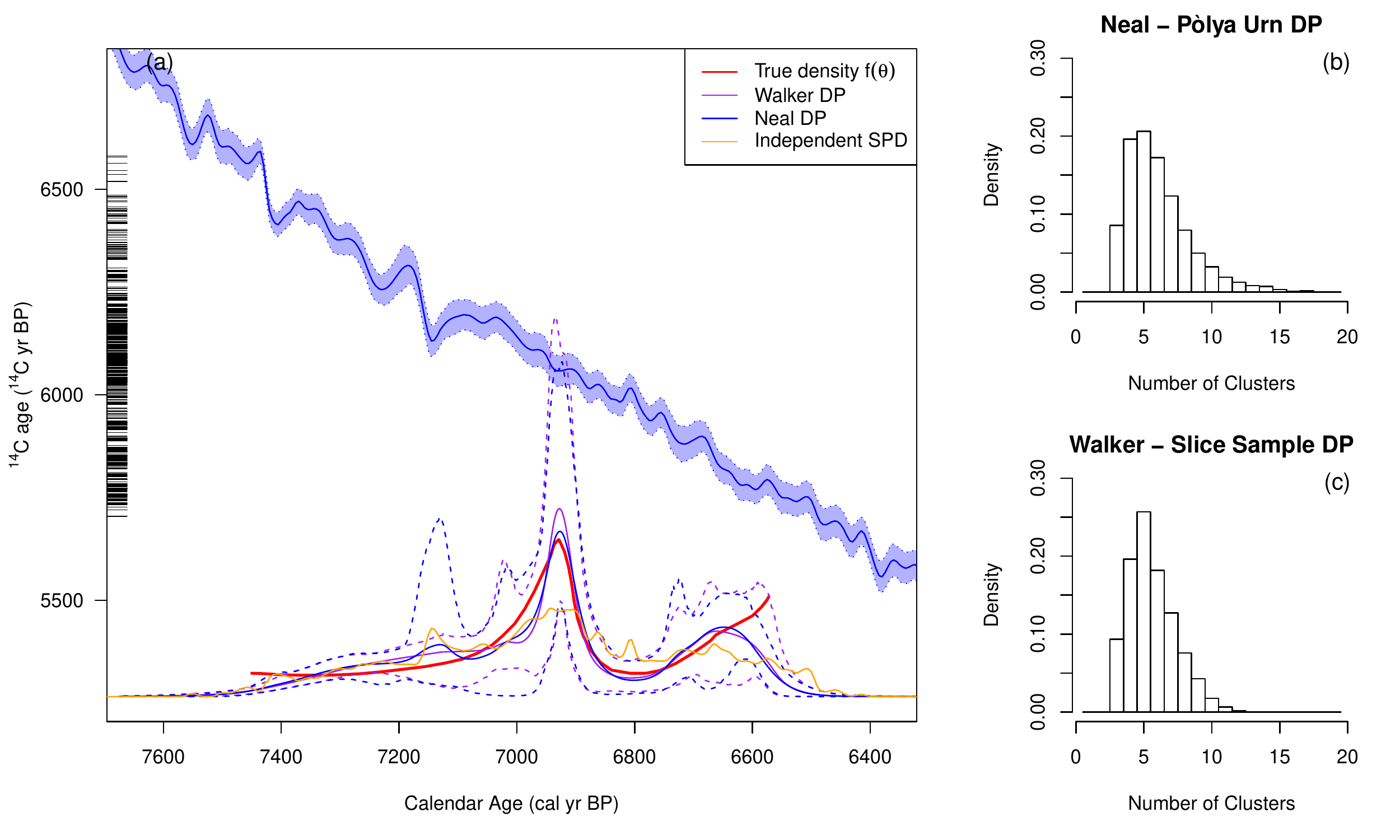}
  \caption{\textit{Estimates of $\hat{f}(\theta)$, based upon summarising 500 \rC observations, when the underlying samples have calendar ages $\theta_i$ that are drawn from a distribution proportional to the estimated population of the Basin of Mexico \citep{McCaa2000} but in a different (older) section of the calibration curve without the Suess effect. Individual panels as for Figure \ref{F:MixtureDensityReconNormal}.} \label{F:ShiftedMixtureDensityReconMexico}}
\end{figure}

To demonstrate that the non-identifiability of the Basin of Mexico population is a consequence of the calibration curve in the time period rather than a failing of the method, and that in other time periods we can accurately reconstruct real-life demography from \rC summarisation, we provide a further example in Figure \ref{F:ShiftedMixtureDensityReconMexico}. Here, to ensure we still have an underlying calendar age density based on real-life, we continue to use the same Basin of Mexico demographic changes \citep[shown in red,][]{McCaa2000} but have shifted the calendar ages back by 6500 cal yrs to a different section of the calibration curve. This calendar age shift means we avoid calibrating during the Suess effect and the unfortunate co-incidence of the narrow population peak with a curve inversion. We can see that here, the DPMM estimates $\hat{f}(\theta)$ are accurately able to reconstruct the underlying demographic changes, albeit with a minor edge effect due to the artificial cutoff in the underlying calendar age density (\textit{c.a.}\ 6600 cal yr BP). The SPD however still fails to reproduce the underlying $f(\theta)$ and, without intervals, is very hard to usefully interpret.    

\subsection{Comparison of  P\'olya Urn and Slice Sampling DP updates}
The simulation study results of Figure \ref{F:SimStudy} and Table \ref{T:SimStudy}, in addition to the sample reconstructions of $f(\theta)$ in Figures \ref{F:MixtureDensityReconNormal}, \ref{F:MixtureDensityReconMexico} and \ref{F:ShiftedMixtureDensityReconMexico}, indicate that the performance of our non-parametric approach is similar whether we use the P\'olya Urn or slice sampling to update the DPMM. The number of distinct clusters to which the $n$ observations are allocated also remains similar between the samplers suggesting that mixing of the P\'olya Urn is still satisfactory. In our implementation, the slice sampling DP updating is considerably faster than the P\'olya Urn. This, in combination with the theoretical improvements that the slice sampling DPMM updates offers to mixing, mean that for our practical examples we only present the estimates using slice sampling DPMM updates. We suggest that the more important application of slice sampling lies in the conditional \rC calibration $\theta_i | x_i, f(\theta)$ given the current cluster allocation.

\section{Practical Examples of \rC Summarisation \label{S:RealExamples}}
We reanalyse three pieces of work where multiple, related, \rC samples have been calibrated and the resultant set of calendar ages summarised and interpreted. All these summaries were originally obtained using SPDs. In all plots, time progresses as we move from left to right, i.e., the older times are shown on the left hand side. Calibration has been performed using IntCal20, the most recent radiocarbon calibration curve \citep{Reimer2020}. In all these examples, the sampler has been run for 50,000 iterations (with the first half discarded) and 5000 samples used for our density estimates and intervals. Since the \rC data in these examples extend over reasonable periods of time, and there are no extremely narrow peaks in our $\hat{f}(\theta)$ estimates with wide probability intervals that coincide with inversions in the curve, we believe the DPMM estimates do enable identifiable inference on the the underlying calendar age density. Some edge effects may however remain if \rC determinations outside certain ranges were removed from the underlying datasets, see Section \ref{S:DensityRecon}.

\subsection{Irish raths}
\begin{figure}[t!]
  \centering
      \includegraphics[width=0.99\textwidth, keepaspectratio]{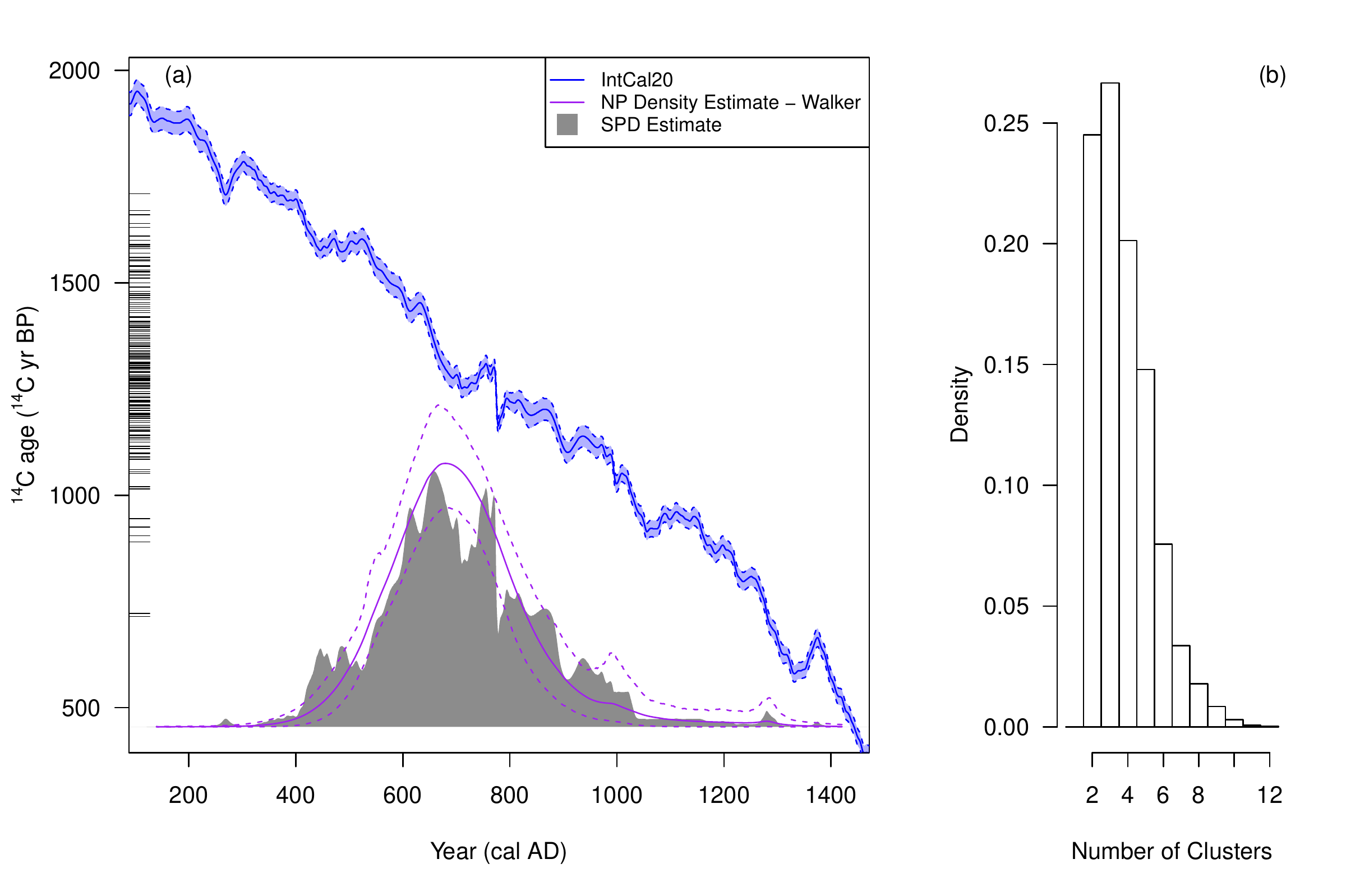}
  \caption{\textit{Prevalence of raths in medieval Ireland based upon 255 \rC determinations \citep{Kerr2014}. Panel (a) Non-parametric Bayesian DPMM estimate (purple, with dashed pointwise 95\% credible intervals) and SPD estimate (shaded grey); Panel (b) the number of distinct DPMM clusters used to model the 255 determinations, the multiple clusters provide the slight positive skew and heavy tails in the predictive density. } \label{F:KerrDensity}}
\end{figure}
Raths, classic Irish medieval farmsteads consisting of a living area surrounded by a bank and ditch, were not used evenly during the medieval period. To investigate their prevalence over time, \citet{Kerr2014} collated 255 \rC determinations relating to their building and use during the early-medieval period (\textit{ca.} AD 400-1150). Figure \ref{F:KerrDensity} presents the SPD obtained from these \rC determinations and our non-parametric Bayes summary estimate. For this \rC data, we had $\textrm{range}(\tilde{\theta}) = 927$ and $\textrm{mad}(\tilde{\theta}) = 132$, giving a lower 5\% quantile of 6 cal yrs for $\sigma_{\textrm{cluster}, j}$, the prior spread of a cluster in our DPMM.

The SPD is highly multimodal, in particular showing two peaks --- initially around AD 660 followed by a decline before a second peak around AD 775. \citet{Kerr2014} warned against interpreting these two peaks with the intervening trough as evidence for a significant underlying change in rath prevalence, believing the trough was a result of the sampling and the structure in the calibration curve. Our Bayesian DPMM summarised density estimate agrees with this view providing a single mode around AD 680. It supports an interpretation that the use of raths increased steadily over time from around AD 400 to its peak around AD 680, before decreasing more slowly as we progressed further into the medieval period. The SPD peak around AD 775 is likely an artefact of the sharp drop in the calibration curve, due to a massive solar proton event \citep{Miyake2012, Heaton2021}, rather than a demographic change. 

\subsection{Irish population decline at the end of the European Iron Age}

\begin{figure}[t!]
  \centering
      \includegraphics[width=0.99\textwidth, keepaspectratio]{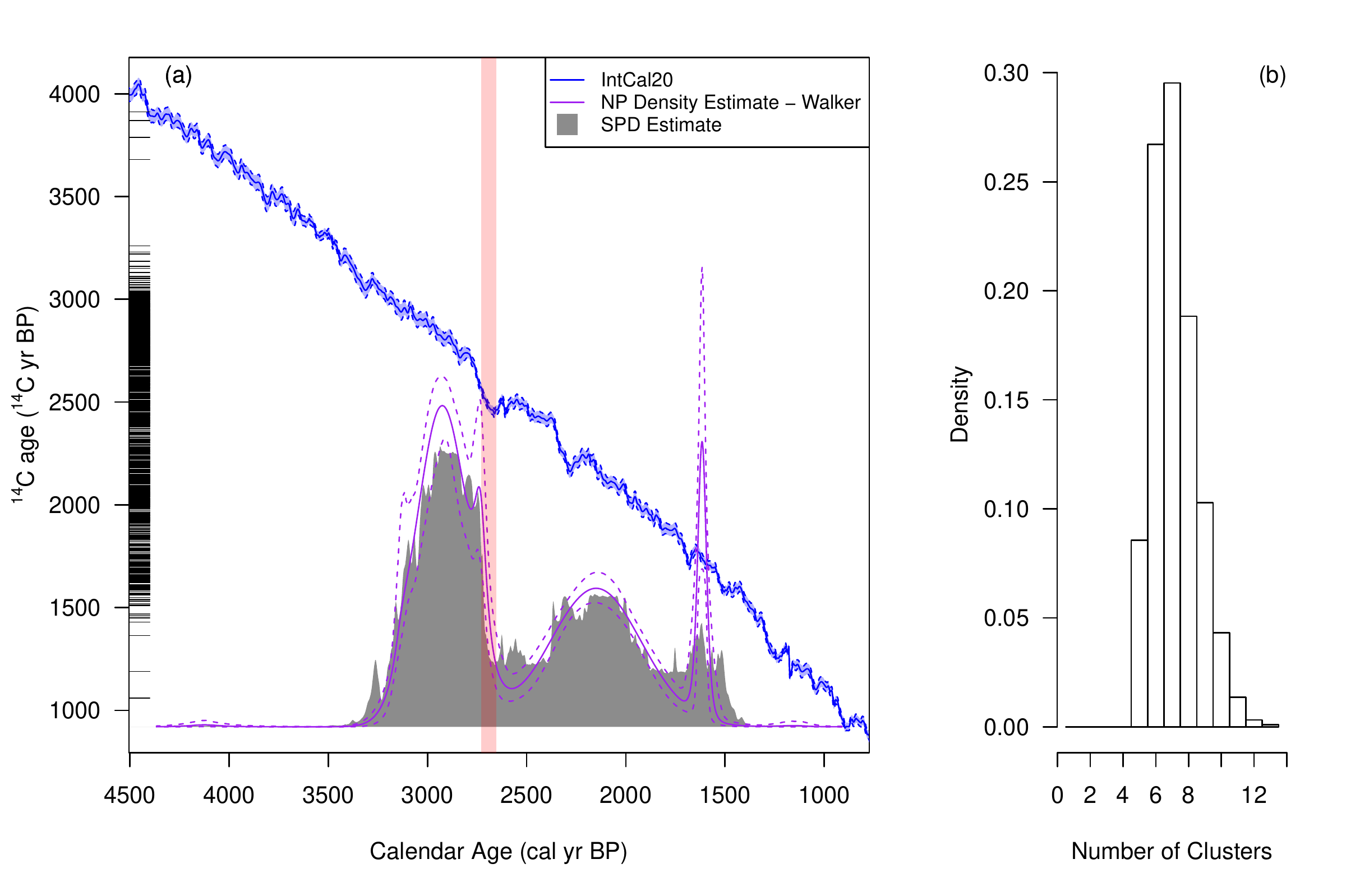}
      
\caption{\textit{Irish population change over time based on summarisation of 2021 \rC determinations \citep{Armit2014}. Panel (a) Non-parametric Bayesian DPMM estimate (purple, with dashed pointwise 95\% credible intervals) and SPD estimate (shaded grey). The red shaded time period shows the range of calendar ages, 2650--2740 cal yr BP, for the believed onset of wetter environment in Ireland. Panel (b) the number of distinct DPMM clusters used to model the 2021 \rC determinations.} \label{F:ArmitDensity}}
\end{figure}

Forecasting societal response and resilience to potential climate change is a pressing global challenge. Crucial insight can be gained by studying our past, which is characterized by rapid environmental changes similar to those we predict for our future. Towards the end of the European Iron Age, we experienced a large deterioration in climate across North-Western Europe. It has been suggested that this led to socio-economic collapses and culture changes across the region. This downturn in climate is proposed to have preceded by a rapid decrease in solar activity around 2800 cal yr BP with a less favourable climate seen across Europe at \textit{ca.} 2700--2750 cal yr BP. In Ireland, this manifested in a much wetter environment \textit{ca.} 2700 cal yr BP \citep{Swindles2007}. To investigate if the onset of this wetter environment led to a population collapse in Ireland, \citet{Armit2014} collated all available \rC determinations from archaeological groups operating within the country. Summarising the calendar age information provided by these 2021 samples (two are removed due to missing \rC measurement uncertainty) provides, through estimation of $f(\theta)$, a proxy for population size over time.

Figure \ref{F:ArmitDensity} shows the SPD and our non-parametric DPMM estimate for $f(\theta)$. For this \rC data, we had $\textrm{range}(\tilde{\theta}) = 3450$ and $\textrm{mad}(\tilde{\theta}) = 525$, giving a lower 5\% quantile of 24 cal yrs for $\sigma_{\textrm{cluster}, j}$, the prior spread of a cluster in our DPMM. While the SPD identifies several main features, a large peak in samples from around 3000 cal yr BP and a smaller peak around 2100 cal yr BP, it is difficult to determine whether the finer scale variations are artefacts of the SPD method or real features of the underlying density $f(\theta)$. In particular, it is unclear precisely when the SPD peak around 3000 cal yr BP begins to drop; and whether the later features around 1550 cal yr BP, are genuine. The DPMM estimate, with credible intervals, provides a clearer interpretation. The density starts to rise from 3400 cal yr BP to a peak at around 2990 cal yr BP after which the density begins to decline. On the decline there appears to be a small secondary peak around 2750 cal yr BP after which the decline is more rapid. Around 2500 cal yr BP the density is low, in agreement with the low volume of evidence for Irish settlements in this period (the early Irish Bronze Age). The density then slowly rises again to suggest increased activity around 2100 cal yr BP before dropping to another low at \textit{ca.} 1670 cal yr BP. This low corresponds to the Irish Dark Ages \citep{CharlesEdwards2000}. A sharp recovery is then indicated by the peak \textit{ca.} 1540 cal yr BP.

Critically, using the DPMM estimate as a proxy for population size, we find that the decline in population (beginning around 2990 cal yr BP) precedes the climate change in Ireland at \textit{ca.} 2700 cal yr BP. We note however the small temporary increase in the predictive density at 2750 cal yr BP which corresponds to a decrease in solar activity. Unlike the SPD, we see that the rise in activity during the 2600--1540 cal yr BP period is smooth; and also that the final peak around 1550 cal yr BP is not an artefact but evidence in support of the sharp recovery in activity corresponding to the end of the Irish Dark Ages as previously identified by \citet{CharlesEdwards2000}.

\subsection{Palaeo-Indian demography}
The extra-terrestrial (ET) impact hypothesis is a highly controversial theory for the cause of the Younger-Dryas (YD), a significant climatic cooling event which lasted from \textit{c.a.} 12,800--11,700 cal yr BP. \citet{Firestone2007} proposed that around 12,900 cal yr BP one or more large ET objects impacted (or exploded) over northern North America with catastrophic consequences including the destabilisation of the Laurentide Ice Sheet leading to the abrupt onset of the YD; the destruction of Pleistocene megafauna; and the collapse of the palaeoindian Clovis population across the North American continent. The hypothesis is significantly disputed, see e.g., \citet{Kennett2009Science, Kennett2009PNAS} for suggested evidence in favour vs. \citet{Pinter2011} for a summary of the argument against. 

\begin{figure}[t!]
  \centering
    \includegraphics[width=0.99\textwidth, keepaspectratio]{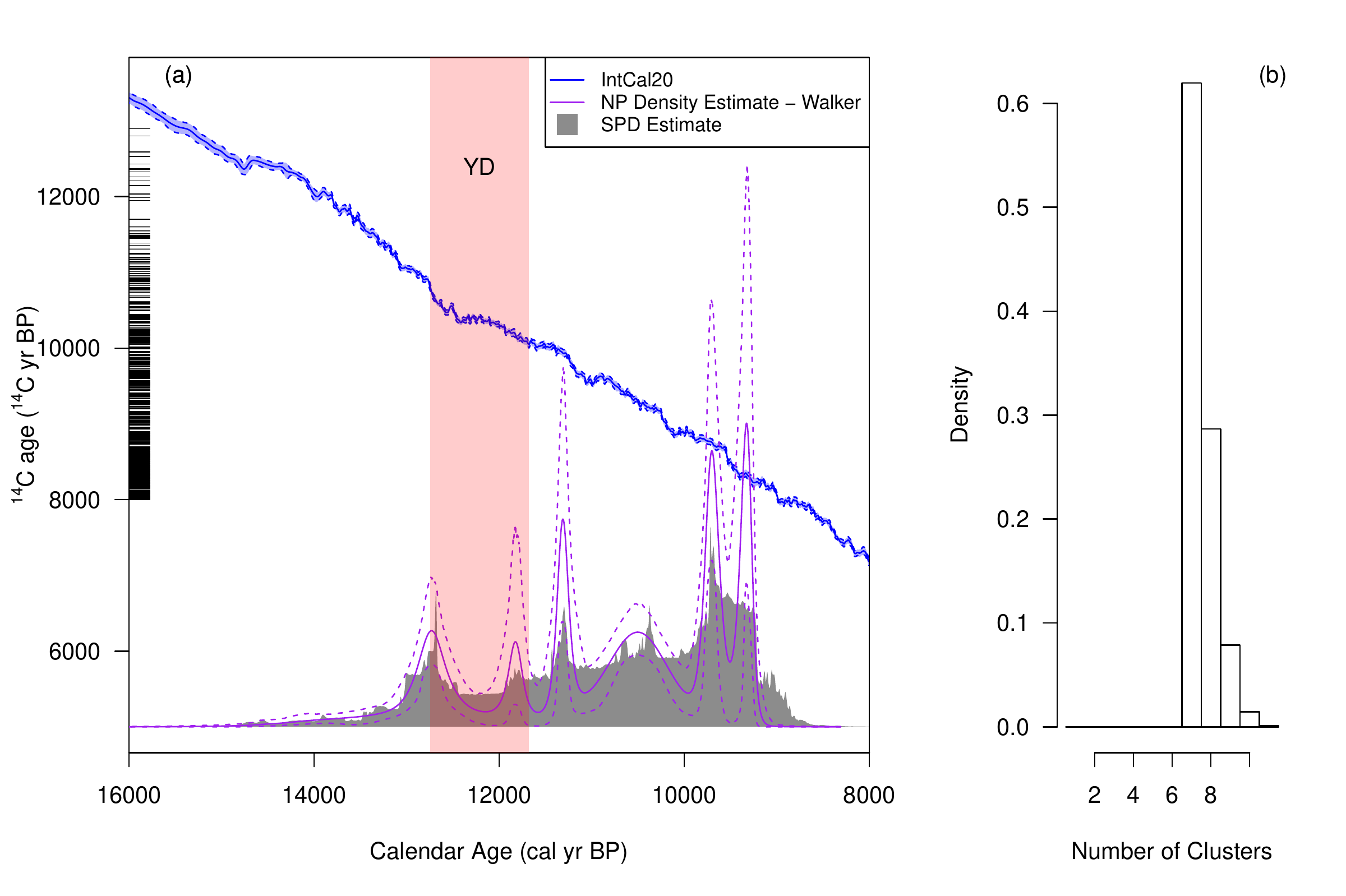} \\
  \caption{\textit{Palaeoindian demography based on summarisation of 628 \rC determinations from distinct N. American archaeological sites \citep{Buchanan2008}. Panel (a) Non-parametric Bayesian DPMM estimate (purple, with dashed pointwise 95\% credible intervals) and SPD estimate (shaded grey). The red shaded time period shows the YD \citep{Rasmussen2014, Reinig2021}. Panel (b) the number of distinct DPMM clusters used to model the 628 \rC determinations.} \label{F:BuchananDensity}}
\end{figure}

One approach taken to assess the plausibility of the ET hypothesis has been to investigate if a collapse in the palaeoindian population is seen directly after the time of the supposed ET impact. \citet{Buchanan2008} collated 628 \rC determinations representing the ages of distinct archaeological sites found across Canada and North America during the time of the palaeoindians. They summarised the calibrated dates of these \rC determinations using SPDs to infer potential changes in population size, concluding that no significant population bottleneck was observed.

In our DPMM reanalysis, we do not intend to argue for, or against, the ET hypothesis. Instead our interest lies in investigating possible broader changes to palaeoindian demography over time. Figure \ref{F:BuchananDensity} presents the SPD of the 628 \rC determinations of \citet{Buchanan2008} together with our non-parametric DPMM density estimate $\hat{f}(\theta)$. For this \rC data, we had $\textrm{range}(\tilde{\theta}) = 6485$ and $\textrm{mad}(\tilde{\theta}) = 1480$, giving a lower 5\% quantile of 67 cal yrs for the prior spread of a cluster in our DPMM. We see five/six distinct peaks in our DPMM density estimate . Since these are well separated in time we can be confident these are not artefacts due to inversions in the calibration curve. The sharp drop at the end of our density on the right hand side around 9000 cal yr BP is a result of Buchanan's site selection criteria --- archaeological sites with \rC determinations younger than 8,000 \rC yr BP were not collated. Moving from oldest to youngest, there appears to be a peak in the number of archaeological sites at 12,800 cal yr BP around the start of the Younger-Dryas. This is followed by a rapid decline in the probability of a site during the YD, except for a potential small peak around 11,800 cal yr BP. We then see a sharp and short lived peak in the calendar age density at 11,250 cal yr BP. This peak corresponds to a few hundred years after the believed end of the YD when average global temperatures would have been rapidly increasing to levels comparable with the present day. The prevalence of sites then appears to decrease again before a more steady increase around 10,250 cal yr BP. There are then two final peaks around 9,500 and 9,100 cal yr BP. Conversely, the SPD approach provides a less clear picture identifying only 4 clear peaks but a much greater amount of noise.

The clear structure in our DPMM density estimate, especially the very sharp peaks, suggest several further questions --- are these sudden peaks of population; do they correspond to potential mass migrations due to climate change; or do they rather relate to the specific archaeological sampling of the collated sites?

\section{Conclusion \label{S:Conclusion}}
Bayesian analyses are now standard in the radiocarbon community, both to create the essential IntCal calibration curve recording the varying proportion of \nC to \rC over time \citep{Reimer2020}; and in the subsequent calibration to convert a \rC determination $x_i$ to a calendar age $\theta_i$. When considering multiple related \rC determinations, where the samples have calendar ages that have been drawn from a underlying, but unknown, wider population $\theta_i \sim f(\theta)$, we obtain better calendar age estimation by calibrating the multiple \rC determinations jointly $\theta_i | x_1, \dots, x_n$ rather than independently $\theta_i | x_i$. By estimating the underlying shared calendar age density $f(\theta)$, we can also summarise the calendar age information provided by the samples and obtain a useful summary of population size or activity over time. 

Standard approaches to joint calibration in the \rC community have required a fixed parametric form for $f(\theta)$. However, in many cases the underlying calendar age distribution from which the samples have been drawn is neither simple nor known in advance and so parametric approaches are not appropriate. Current non-parametric approaches such as SPDs \citep{Williams12} lack statistical rigour and a theoretical underpinning in keeping with the Bayesian calibration framework. 

In this paper, we develop a non-parametric latent DPMM approach which is rigorous and provides a fully Bayesian scheme. We simultaneously estimate both the individual calendar ages $\theta_i | x_1, \dots, x_n$, for $i = 1, \ldots, n$; and the underlying shared calendar age density $f(\theta)$. We present a simulation study indicating that our joint DPMM-based approach to calibration offers 15--30\% improvements in estimation of the calendar ages $\hat{\theta_i}$ compared with independent calibration. We also show how the DPMM provides useful estimates of the underlying shared calendar age density $\hat{f}(\theta)$, both in our simulation examples and in practical settings.

Future work could consider alternative distributions, beyond the normals used here, as the components in the infinite mixture used to model the underlying $f(\theta)$. Distributions designed to match beliefs about the true mixture of phases in a study could provide additional insight into the number of distinct activity periods over time; and further improve both the accuracy of the calibration and the estimation of $f(\theta)$. Further valuable extensions might ensure the approach is robust to potential outliers in the \rC determinations $x_i$, or non-normal measurement uncertainty; and incorporate the covariance information present in the calibration curve.

Finally, we remind users that taphonomic loss often occurs in archaeological and radiocarbon contexts, whereby we are more likely to radiocarbon date objects from certain time periods due to differences in both sampling and survival \citep{Contreras2014}. Where we believe taphonomic loss to be significant, or sampling to be non-random, caution should be exercised in directly interpreting our summarised calendar age density estimates as representative of underlying population activity.

\section*{Code and Data Availability}
Code to implement the proposed methods, and to reproduce all the work in this paper, including the examples and the simulation study, is available via GitHub (\url{https://github.com/TJHeaton/NonparametricCalibration}). Coding is performed in \textsf{R} \citep{RLang}.

\section*{Acknowledgements}
This work was supported by a Leverhulme Trust Fellowship RF-2019-140\textbackslash 9. I thank Christopher Bronk Ramsey for introducing me to the challenge of calibrating and summarising multiple \rC determinations, as well as helpful discussions in developing the described approach. I am also grateful to Paul G.\ Blackwell and Marie-Anne Vibet for their comments, as well as the IntCal group for the time they have spent answering my many questions. Finally thanks go to the referees and Prof.\ Richard Boys for their valuable suggestions that have significantly improved the work.

\bibliographystyle{elsarticle-harv}
\bibliography{NPRefsShort}

\clearpage
\appendix
\renewcommand\thetable{\Alph{section}\arabic{table}}
\renewcommand\thefigure{\Alph{section}\arabic{figure}}
\setcounter{figure}{0}
\setcounter{table}{0}

\section{Simulation Study}
% latex table generated in R 3.4.4 by xtable 1.8-2 package
% Sat Jun 15 17:05:17 2019
\begin{sidewaystable}[h]
 \centering
\begin{tabular}{||r|r|r|r|r||r|r|r|r||r|r|r|r||r|r|r|r||}
  \hline
   & \multicolumn{16}{|c||}{Example 1: Single Underlying Normal Phase} \\
  & \multicolumn{16}{|c||}{($\mu_\phi = 10000$, $\theta_i \in [100, 49500]$)} \\
   & \multicolumn{8}{|c||}{Neal P\'olya Urn} &  \multicolumn{8}{|c||}{Walker slice DPMM}  \\
  & \multicolumn{4}{|c||}{$l_1$-loss} &  \multicolumn{4}{|c||}{$l_2$-loss} & \multicolumn{4}{|c||}{$l_1$-loss} &  \multicolumn{4}{|c||}{$l_2$-loss} \\
 & \multicolumn{1}{|c|}{Prop. imp} & \multicolumn{3}{|c||}{\% Improvement} & \multicolumn{1}{|c|}{Prop. imp} & \multicolumn{3}{|c||}{\% Improvement} & \multicolumn{1}{|c|}{Prop. imp} & \multicolumn{3}{|c||}{\% Improvement} & \multicolumn{1}{|c|}{Prop. imp} & \multicolumn{3}{|c||}{\% Improvement} \\
 n & (95\% CI)  & Mean & Max & Min & (95\% CI) & Mean & Max & Min & (95\% CI)  & Mean & Max & Min & (95\% CI) & Mean & Max & Min \\ \hline
50 & 96 (86,100) & 25 & 64.9 & -9.4 & 98 (89,100) & 41.6 & 87 & -11.4 & 100 (93,100) & 25.2 & 65.5 & 0.1 & 100 (93,100) & 41.9 & 89.1 & 1.7 \\ 
  100 & 98 (89,100) & 20.6 & 62.6 & -0.5 & 98 (89,100) & 34.4 & 85.1 & -1.3 & 98 (89,100) & 20.3 & 65 & -1.1 & 98 (89,100) & 33.8 & 89.2 & -1.9 \\ 
  200 & 98 (89,100) & 17.3 & 47.4 & -12.9 & 96 (86,100) & 29.8 & 71.2 & -23.8 & 98 (89,100) & 17.3 & 49.5 & -1.7 & 96 (86,100) & 29.6 & 71.7 & -5.2 \\ 
  500 & 100 (93,100) & 21.2 & 58.8 & 0.3 & 100 (93,100) & 35.4 & 83.5 & 0.6 & 100 (93,100) & 20.7 & 67.8 & 0.3 & 98 (89,100) & 34.4 & 87.7 & -0.3 \\  
   & \multicolumn{16}{|c||}{}\\
  & \multicolumn{16}{|c||}{Example 2: Three Underlying Normal Phases} \\
   & \multicolumn{16}{|c||}{($\mu_\phi = 3000$, $\theta_i \in [100, 15000]$)} \\
  & \multicolumn{8}{|c||}{Neal P\'olya Urn} &  \multicolumn{8}{|c||}{Walker slice DPMM}  \\
  & \multicolumn{4}{|c||}{$l_1$-loss} &  \multicolumn{4}{|c||}{$l_2$-loss} & \multicolumn{4}{|c||}{$l_1$-loss} &  \multicolumn{4}{|c||}{$l_2$-loss} \\
 & \multicolumn{1}{|c|}{Prop. imp} & \multicolumn{3}{|c||}{\% Improvement} & \multicolumn{1}{|c|}{Prop. imp} & \multicolumn{3}{|c||}{\% Improvement} & \multicolumn{1}{|c|}{Prop. imp} & \multicolumn{3}{|c||}{\% Improvement} & \multicolumn{1}{|c|}{Prop. imp} & \multicolumn{3}{|c||}{\% Improvement} \\
 n & (95\% CI)  & Mean & Max & Min & (95\% CI) & Mean & Max & Min & (95\% CI)  & Mean & Max & Min & (95\% CI) & Mean & Max & Min \\ \hline
50 & 94 (83,99) & 9.1 & 31.1 & -5.5 & 88 (76,95) & 16.2 & 54.4 & -10 & 92 (81,98) & 9.3 & 32.9 & -5.1 & 86 (73,94) & 16.6 & 54.9 & -9.3 \\ 
  100 & 100 (93,100) & 11.7 & 44.4 & 1.8 & 100 (93,100) & 21 & 69.9 & 3.1 & 100 (93,100) & 11.4 & 47 & 0.9 & 100 (93,100) & 20.3 & 72.5 & 3.2 \\ 
  200 & 98 (89,100) & 10.9 & 47 & -0.7 & 98 (89,100) & 19.6 & 71.4 & -4.5 & 98 (89,100) & 11 & 46.8 & -0.5 & 98 (89,100) & 19.9 & 71.3 & -3.8 \\ 
  500 & 100 (93,100) & 12.5 & 41.5 & 1.7 & 100 (93,100) & 22.6 & 65.5 & 3.5 & 100 (93,100) & 12 & 35.3 & 0.5 & 100 (93,100) & 21.9 & 58.6 & 2.6 \\  
  \hline 
  & \multicolumn{16}{|c||}{}\\
  & \multicolumn{16}{|c||}{Example 3: Uniform Distribution} \\
  & \multicolumn{16}{|c||}{($\theta_i \in [100, 15000]$)} \\
   & \multicolumn{8}{|c||}{Neal P\'olya Urn} &  \multicolumn{8}{|c||}{Walker slice DPMM}  \\
  & \multicolumn{4}{|c||}{$l_1$-loss} &  \multicolumn{4}{|c||}{$l_2$-loss} & \multicolumn{4}{|c||}{$l_1$-loss} &  \multicolumn{4}{|c||}{$l_2$-loss} \\
 & \multicolumn{1}{|c|}{Prop. imp} & \multicolumn{3}{|c||}{\% Improvement} & \multicolumn{1}{|c|}{Prop. imp} & \multicolumn{3}{|c||}{\% Improvement} & \multicolumn{1}{|c|}{Prop. imp} & \multicolumn{3}{|c||}{\% Improvement} & \multicolumn{1}{|c|}{Prop. imp} & \multicolumn{3}{|c||}{\% Improvement} \\
 n & (95\% CI)  & Mean & Max & Min & (95\% CI) & Mean & Max & Min & (95\% CI)  & Mean & Max & Min & (95\% CI) & Mean & Max & Min \\ \hline
50 & 96 (86,100) & 18.9 & 63.9 & -3.1 & 92 (81,98) & 32 & 86.9 & -5.6 & 94 (83,99) & 19.2 & 59.3 & -4.1 & 94 (83,99) & 32.5 & 82.3 & -8.6 \\ 
  100 & 96 (86,100) & 14.8 & 74.2 & -60.8 & 96 (86,100) & 24.3 & 91.8 & -87.3 & 90 (78,97) & 14.6 & 75.5 & -57.8 & 92 (81,98) & 23 & 92.9 & -82.6 \\ 
  200 & 96 (86,100) & 19 & 70.1 & -34.5 & 96 (86,100) & 32 & 89.7 & -28.9 & 96 (86,100) & 19.8 & 71.5 & -10.6 & 100 (93,100) & 33.4 & 89.5 & 0.1 \\ 
  500 & 92 (81,98) & 17.3 & 59.4 & -13.4 & 84 (71,93) & 28.2 & 82.2 & -30.8 & 96 (86,100) & 16.7 & 53 & -2.8 & 92 (81,98) & 28.1 & 76.4 & -8.1 \\ 
   \hline
\end{tabular} 
\caption{Simulation study showing the improvement in calendar age estimation of $\theta_i$ achieved when using our joint non-parametric approach, compared to independent calibration of each \rC determination, if the calendar ages of the samples are known to be related. The first column gives the proportion of (the 50) runs which provide an overall improvement in loss over independent calibration. The accompanying CIs were obtained using the method of \citet{Clopper1934}. The later columns the mean, maximum, and minimum improvement seen in the 50 simulation runs. \label{T:SimStudy}}   
\end{sidewaystable}  

\clearpage

\section{Reconstructing a Uniform Calendar Age Density \label{S:AppendixUnifRecon}}
\begin{figure}[h]
  \centering
      \includegraphics[width=0.99\textwidth, keepaspectratio]{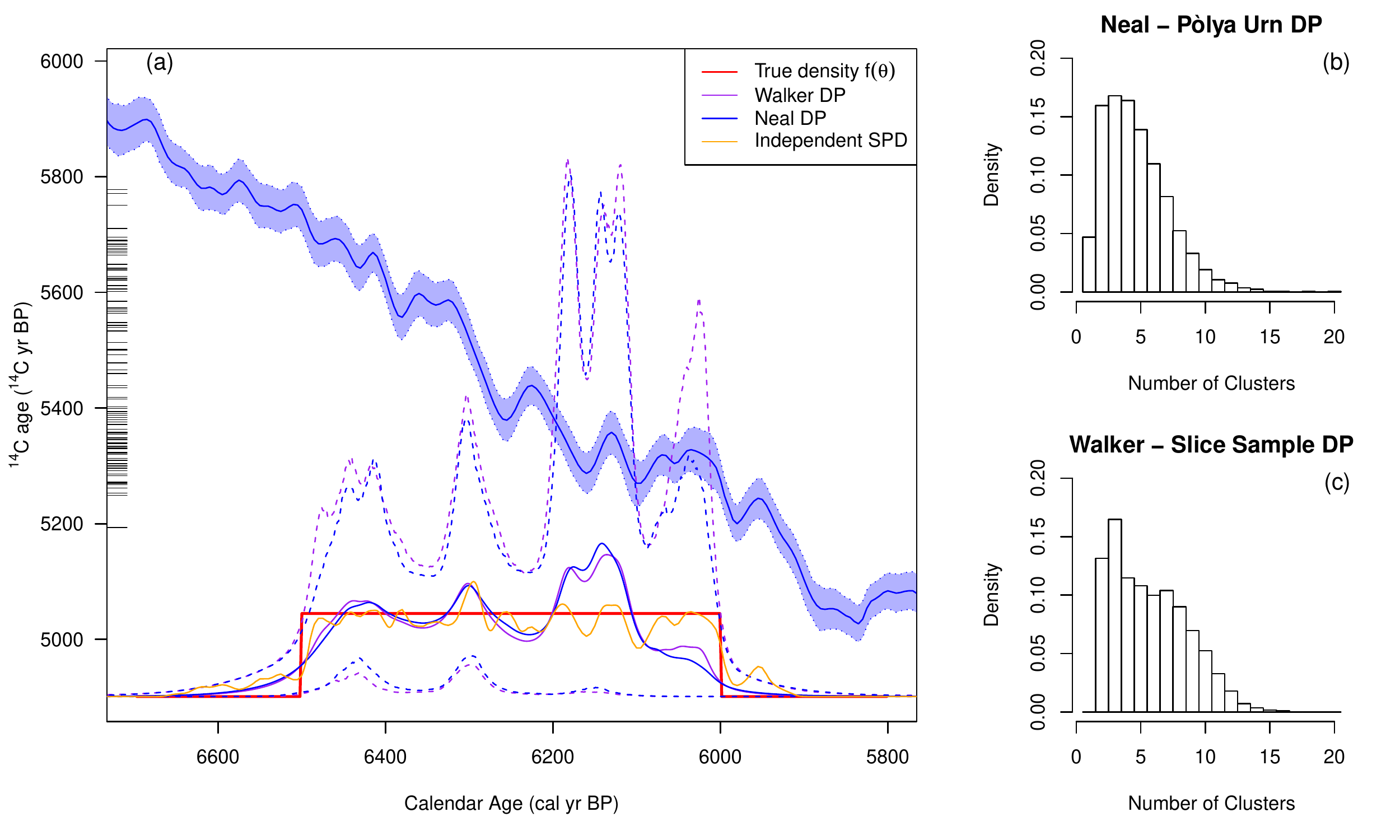}
  \caption{\textit{Estimates of $\hat{f}(\theta)$, based upon summarising 100 \rC observations, when the underlying samples have calendar ages drawn from a uniform distribution: $\theta_i \sim f(\theta) = \textrm{U}[6000, 6500]$ cal yrs, and corresponding \rC determinations $X_i \sim N(m(\theta_i), \sigma_\textrm{obs}^2 + \rho(\theta_i)^2)$ \rCyrs. Individual panels as for Figure \ref{F:MixtureDensityReconNormal}.} \label{F:MixtureDensityReconUniform}}
\end{figure}

Figures \ref{F:MixtureDensityReconUniform} show an illustrative DPMM estimate $\hat{f}(\theta)$ for the summarised calendar age density based upon 100 \rC determinations for which the underlying calendar ages were drawn from a uniform distribution. We compare the true underlying calendar age density (shown in red) against the estimates obtained via our non-parametric Bayes, blue (P\'olya Urn DP updating) and purple (slice sampling); and the standard SPD estimate. The pointwise mean for the predictive density is shown with a solid line; 95\% pointwise credible intervals, based upon the individual realisations, are shown with dashed lines.   

None of the methods perform well in reconstructing $f(\theta)$. Both the P\'olya Urn and slice sampling DPMM updates locate the shared density correctly but the use of normal distributions as the components in the DPMM result in estimates that do not have the sharp drop outside the support of the underlying uniform and generate spurious peaks in the centre of the density reconstruction. Note however that the pointwise credible intervals are wide suggesting the model is struggling to fit properly. This is to be expected since reconstructing a uniform using a mixture of normal distributions is extremely challenging. Interestingly, both DPMM approaches tend to model the 100 \rC determinations as belonging to a small number of clusters, see panels (b) and (c). Here the independent SPD approach perhaps offers better reconstruction of $f(\theta)$ although, as shown in Section \ref{S:SimStudyResults},  the estimation of the individual $\theta_i$ remains significantly worse since the SPD approach does not use the estimate $\hat{f}(\theta)$ in calibrating each determination $x_i$. Importantly, we would expect DPMM performance to be much improved should we use mixture components in our DPMM that better resemble the underlying $f(\theta)$.

\end{document}